# 'Rainbows' in homogeneous and radially inhomogeneous spheres: connections with ray, wave and potential scattering theory

John A. Adam
Department of Mathematics & Statistics
Old Dominion University
Norfolk, VA 23508

**Abstract:** This chapter represents an attempt to summarize some of the direct and indirect connections that exist between ray theory, wave theory and potential scattering theory. Such connections have been noted in the past, and have been exploited to some degree, but in the opinion of this author, there is much more yet to be pursued in this regard. This article provides the framework for more detailed analysis in the future. In order to gain a better appreciation for a topic, it is frequently of value to examine it from as many complementary levels of description as possible, and that is the objective here. Drawing in part on the work of Nussenzveig, Lock, Debye and others, the mathematical nature of the rainbow is discussed from several perspectives. The primary bow is the lowest-order bow that can occur by scattering from a spherical drop with constant refractive index $n$, but zero-order (or direct transmission) bows can exist when the sphere is radially inhomogeneous. The refractive index profile automatically defines a scattering potential, but with a significant difference compared to the standard quantum mechanical form: the potential is $k$-dependent. A consequence of this is that there are no bound states for this system. The correspondences between the resonant modes in scattering by a potential of the 'well-barrier' type and the behavior of electromagnetic 'rays' in a transparent (or dielectric) sphere are discussed. The poles and saddle points of the associated scattering matrix have quite profound connections to electromagnetic tunneling, resonances and 'rainbows' arising within and from the sphere. The links between the various mathematical and physical viewpoints are most easily appreciated in the case of constant $n$, thus providing insight into possible extensions to these descriptions for bows of arbitrary order in radially inhomogeneous spheres (and cylinders).

## 1. Introduction: the rainbow – its scientific and mathematical beauty

"Rainbows have long been a source of inspiration both for those who would prefer to treat them impressionistically or mathematically. The attraction to this phenomenon of Descartes, Newton, and Young, among others, has resulted in the formulation and testing of some of the most fundamental principles of mathematical physics."
K. Sassen [1]

"The rainbow is a bridge between two cultures: poets and scientists alike have long been challenged to describe it...Some of the most powerful tools of mathematical physics were devised explicitly to deal with the problem of the rainbow and with closely related problems. Indeed, the rainbow has served as a touchstone for testing theories of optics. With the more successful of those theories it is now possible to describe the rainbow mathematically, that is, to predict the distribution of light in the sky. The same methods can also be applied to related phenomena, such as the bright ring of color called the glory, and even to other kinds of rainbows, such as atomic and nuclear ones."

H.M. Nussenzveig [2]

"The theory of the rainbow has been formulated at many levels of sophistication. In the geometrical-optics theory of Descartes, a rainbow occurs when the angle of the light rays emerging from a water droplet after a number of internal reflections reaches an extremum. In Airy's wave-optics theory, the distortion of the wave front of the incident light produced by the internal reflections describes the production of the supernumerary bows and predicts a shift of a few tenths of a degree in the angular position of the rainbow from its geometrical-optics location. In Mie theory, the rainbow appears as a strong enhancement in the electric field scattered by the water droplet. Although the Mie electric field is the exact solution to the light-scattering problem, it takes the form of an infinite series of partial wave contributions that is slowly convergent and whose terms have a mathematically complicated form. In the complex angular momentum theory, the sum over partial waves is replaced by an integral, and the rainbow appears as a confluence of saddle-point contributions in the portion of the integral that describes light rays that have undergone *m\** internal reflections within the water droplet."

J. A. Lock [3]

\* In this chapter, $p-1$ will replace *m*, where $p \geq 1$.

**1.1 Complementary domains of description**

This chapter addresses three related topics: the existence of direct transmission (or zero-order) bows in radially inhomogeneous spheres, the Mie solution of electromagnetic scattering, and the associated wave-theoretic/potential scattering connection, to be discussed in detail below. This connection is well illustrated in a series of recent papers by Lock [4-6] (see Section 3).

Geometrical optics and wave (or physical) optics are two very different but complementary approaches to describing many optical phenomena, and here, specifically, the rainbow. However, there is a broad 'middle ground', the '*semiclassical*' régime. Thus there are essentially three domains within which scattering phenomena may be described: the scattering of waves by objects which in size are (i) small, (ii) comparable with, and (iii) large, compared to the wavelength of the incident (plane wave) radiation. There may be considerable overlap of region (ii) with the others, depending on the problem of interest, but basically, the wave-theoretic principles in region (i) tell us why the sky is blue (amongst many other things!). At the other extreme, the 'classical' domain (iii) enables us in particular to be able to describe the basic features of the rainbow in terms of ray optics. The wave-particle duality so fundamental in quantum mechanics is relevant to region (ii) because the more subtle features exhibited by such phenomena involve both these aspects of description and explanation. Indeed, it is useful to relate (somewhat loosely) the régimes (i) - (iii) above to three domains, as stated by Grandy [7]:

(a) *The classical domain*: geometrical optics; particle and particle/ray-like trajectories.

(b) *The wave domain*: physical optics; acoustic and electromagnetic waves; quantum mechanics.

(c) *The semiclassical domain*: "the vast intermediate region between the above two, containing many interesting physical phenomena."

Geometrical optics is associated with 'real' rays, but their analytic continuation to complex values of some associated parameters enables the concept of 'complex rays' to be used, often in connection with surface or 'evanescent' rays travelling along a boundary while penetrating the less dense medium in an exponentially damped manner. However, complex rays can also be used to describe the phenomenon of *diffraction*: the penetration of light into regions that are forbidden to the real rays of geometrical optics [8], so there are several different contexts in which this term can be used. In fact, the primary bow light/shadow transition region is associated physically with the confluence of a pair of geometrical rays and their transformation into complex rays; mathematically this corresponds to a pair of real saddle points merging into a complex saddle point. For the primary bow then, the two (supernumerary) rays coalesce when they are incident on the sphere surface at the Descartes angle, and the subsequent vanishing of these rays is associated with the complex ray on the shadow side of the rainbow. This does not involve 'grazing incidence' at all. On the other hand, rays that graze the sphere *and* just miss grazing it may 'tunnel' into the interior, or more accurately, *both* of these regions together form an 'edge region' that gives rise to the tunneling ray.. This phenomenon is well-known in quantum mechanics; specifically, tunneling through a classically-forbidden potential barrier. Because it occurs in the edge region of semiclassical scattering, it permits grazing rays (and those just outside the sphere) to interact with it (and contribute to the radiation field) [8-10]. As shown by Nussenzveig in a series of very elegant but technical papers [9-12], scattering of scalar waves by a transparent sphere is in many respects isomorphic to the problem of scattering of particles by a spherical potential well. In quantum mechanics, as will be shown later in this chapter, the bound states of a potential well correspond to poles in the elements of a certain matrix, the *scattering matrix*, on the negative real energy axis, whereas *resonances* of the well (as we shall see) correspond to poles that are just below the positive real energy axis of the second Riemann sheet associated with those matrix elements. The closer these poles are to the real axis, the more the resonances behave like very long-lived bound states, or 'almost bound' states of the system. In very simplistic terms, if a particle with a resonance energy is 'shot' at the well from far enough away, it is captured by the well for a considerable time, and acts like a bound particle, but eventually it escapes from the well (this, for example, is a crude description of the mechanism of $\alpha$-decay from a nucleus, though that is a decay phenomenon, not a scattering one). The reciprocal of the half-width of the resonance is a measure of the lifetime of the resonance particle in the well.

In view of all this then, mathematically at least, a primary 'rainbow' is, amongst other things [13, 14]:

(1) a concentration of light rays corresponding to an extremum of the deviation or scattering angle (this extremum is identified as the Descartes' or rainbow ray); (2) a caustic, separating a two-ray region from a 0-ray (or shadow) region; (3) an integral superposition of waves over a (locally) cubic wave-front (the *Airy approximation*); (4) a coalescence of two real saddle-points; (5) a result of scattering by a square-well potential; (6) an example of 'Regge-pole dominance', and (7) a *fold diffraction catastrophe*. Most of these complementary descriptions will not be discussed here; instead the reader is referred to references [13, 14] for further details.

## 2. Scattering by a transparent sphere: ray description

In the following discussion, *i* refers to the angle of incidence for the incoming ray, *r* is the radial distance within a sphere of radius *a* (which may be taken to be unity) and $D(i)$ is the deviation undergone by the ray from its original direction. Below, the subscripts *0* and *1* will be used to distinguish the respective deviations of the exiting ray for the direct transmission (or zero-order) and the primary bow. For $p-1$ internal reflections in a spherical droplet of *constant* refractive index $n > 1$, straightforward geometrical optics reveals that the deviation from its original direction of a ray incident from infinity upon the sphere at angle of incidence *i* is, in radians ($i \in [0, \pi/2]$)

$$D_{p-1}(i) = (p-1)\pi + 2i - 2p \arcsin\left(\frac{\sin i}{n}\right). \quad (1)$$

In general, an extremum of this angle exists at $i = i_c$, where

$$i_c = \arccos\left[\frac{n^2 - 1}{p^2 - 1}\right]^{1/2}, \quad p > 1. \qquad (2)$$

Naturally, for real optical phenomena such as rainbows, $n$ is such that $i_c$ exists. A primary bow corresponds to $p = 2,$ a secondary bow to $p = 3,$ and so forth. That a zero-order (or direct transmission bow) corresponding to $p = 1$ cannot exist for constant $n$ is readily shown from equation (1). Nevertheless, it has been established that such relative extrema (for zero and higher-order bows) can exist for radially inhomogeneous spheres (see [15, 16] for more details). In fact, multiple zero order and primary bows may exist depending on the refractive index profile. A well-known result is that the curvature of the ray path is towards regions of higher refractive index $n$. This is a consequence of Snel's law of refraction generalized to continuously varying media. Thus within the sphere, if $dn(r)/dr \equiv n'(r) < 0$ an incoming ray bends towards the origin; if $n'(r) > 0,$ it bends away from it. From Figure 1 it can be seen that for direct transmission in the former case,

(*Figure 1 near here*)

$$i + 2\Theta(i) + (i - |D_0(i)|) = \pi \Rightarrow |D_0(i)| = 2i - \pi + 2\Theta(i). \qquad (3)$$

In this equation, $2\Theta(i)$ is the angle through which the radius vector turns from the point at which the ray enters the sphere to its point of exit. It is readily noted that for one internal reflection (corresponding to a primary bow)

$$|D_1(i)| = 2i - \pi + 4\Theta(i). \qquad (4)$$

In what follows the absolute value notation will be dropped. The deviation formulae can be extended to higher order bows in an obvious fashion. The quantity $\Theta(i)$ is an improper definite integral to be defined in Section 2.1. Analytic expressions for $\Theta(i)$ are difficult to obtain except for a few specific $n(r)$ profiles; several examples are indicated below. For a constant refractive index $\Theta(i)$ is a standard integral resulting in the inverse secant function, and can be readily evaluated. Specifically,

$$D_0(i) = 2i - 2\tilde{r}(i) \text{ and } D_1(i) = 2i + \pi - 4\tilde{r}(i), \qquad (5a, 5b)$$

where $\tilde{r}(i)$ is the angle of refraction inside the sphere. Of course, these results are readily determined from elementary geometry and are the $p = 1$ and $p = 2$ cases referred to earlier. As already noted, there can be no 'zero-order rainbow' for the direct transmission of sunlight in uniform spheres, only primary and secondary bows (ignoring theoretically possible but practically almost unobservable higher-order bows).

(*Figure 2 near here*)

In Figure 2 the dashed curve $D_h$ represents the deviation $D_1(i)$ through a homogeneous sphere of *constant* refractive index $n = 4/3$. The other graphs represent the deviations corresponding to a zero bow and a primary bow for the particular (but arbitrary) choice of refractive index

$$n_1(r) = 1.3 - 0.2\cos\{[1.9(r - 0.85)]^2\}. \quad (6)$$

Note that both $D_0(i)$ and $D_1(i)$ exhibit fairly broad double extrema in this case. It is interesting to note that the relative maximum for $D_1$ is much less pronounced than that for $D_0$. Further discussion of such extrema can be found in [16].

**2.1 The ray path integral**

In a spherically symmetric medium with refractive index $n(r)$ each ray path satisfies the following equation [35]

$$rn(r)\sin\phi = \text{constant}, \quad (7)$$

where $\phi$ is the angle between the radius vector $r$ and the tangent to the ray at that point (note that $r = |r|$). This expression may be thought of as the optical analogue of the conservation of angular momentum for a particle moving under the action of a central force. The result, known as *Bouguer's formula* (for Pierre Bouguer, 1698-1758), implies that all the ray paths $r(\theta)$ are curves lying in planes through the origin ($\theta$ is the polar angle). Elementary differential geometry establishes that

$$\sin\phi = \frac{r(\theta)}{\sqrt{r^2(\theta) + (dr/d\theta)^2}}. \quad (8)$$

From this the angular deviation of a ray $\Theta(i)$ within the sphere can be determined and subsequently the total angle of deviation $D(i)$ through which an incoming ray at angle of incidence $i$ is rotated. From this the formula for $\Theta(i)$ is found to be

$$\Theta(i) = \sin i \int_{r_c(i)}^{1} \frac{dr}{r\sqrt{r^2 n^2(r) - \sin^2 i}}. \quad (9)$$

The lower limit $r_c(i)$ is the point at which the integrand is singular and is therefore the solution of equation (10) below in which (for a *unit* sphere), $\sin i$ is the *impact parameter*. The quantity $r_c(i)$ is the radial point of closest approach to the center of the sphere, sometimes called the *turning point*. The value of $r_c(i)$ is determined implicitly from the following expression

$$\eta(r_c(i)) \equiv r_c(i) n(r_c(i)) = \sin i. \quad (10)$$

The nature of $\eta(r) = rn(r)$ will be very significant in what follows; in particular, $r_c(i)$ will have only one value if $\eta(r)$ is a monotonic function. The integral in equation (9) can be evaluated analytically in certain special cases. Consider first the (somewhat unphysical and singular) power law profile $n(r) = n(R)(r/R)^m$ where $m$ can be of either sign [17]. By a judicious change of variable this can be reduced to the standard result for a constant refractive index. For the choice of a 'shifted hyperbolic' profile of the form $n(r) = (ar+b)^{-1}$ the integral (9) can be evaluated in terms of elementary transcendental functions [15]. The complexity of these integrals increases rapidly with even relatively simple expressions for $n(r)$. In the case of a linear profile, equation (3) can be evaluated in terms of incomplete elliptic integrals of the first and third kinds [18, 19]. A parabolic profile of the form $n(r) = a - br^2$ also yields a result also in terms of a purely imaginary elliptic integral of the third kind [19].

Whether the ray path integral is evaluated analytically or numerically, it contributes to the *direct problem* of geometrical optics, namely (for direct transmission) the total angular deviation $2\Theta(i)$ of the ray inside the sphere for a given profile $n(r)$. Coupled with the refraction at the (in general discontinuous) boundary entrance and exit points this naturally yields the total deviation of an incoming ray as a function of its angle of incidence. The corresponding *inverse problem* is to determine the profile $n(r)$ from knowledge of the observable deflection function $D(i)$ (note that $D(i) = D(\Theta(i))$. This is generally more difficult to accomplish. Another reason for pursuing the inverse problem is that it would be valuable to find at least some sufficient conditions under which inhomogeneous spheres can exhibit bows of any order, but especially of zero order (particularly with regard to industrial techniques such as rainbow refractometry, for example; see references in [16] ). By choosing a generic profile for $D_0(i)$ or $D_1(i)$ for example, it should be possible in principle to examine the implications on $n(r)$ for such profiles. From a strict mathematical point of view, inverse problems in general are notorious for their lack of solution uniqueness. In practical terms it is not significant in this context, and we shall address the topic no further here.

## 2.2 Properties of $\eta(r)$ and interpretation of the ray path integral

A careful analysis of the integral (9) for $\Theta(i)$ in the neighborhood of the singularity yields two possibilities depending on whether or not $\eta(r)$ is a monotone increasing function:

(i) <u>Monotonic case</u>. If $\eta'(r_c) \neq 0$, then in the neighborhood of $r = r_c$, the integral for $\Theta$ has the dominant behavior $(r - r_c)^{1/2}$ which tends to zero as $r \to r_c^+$.
(ii) <u>Non-monotonic case</u>. If $\eta'(r_c) = 0$, then in the neighborhood of $r = r_c$, the integral for $\Theta$ has the dominant behavior $\ln|r - r_c|$ which tends to $-\infty$ as $r \to r_c^+$.

To see this, we expand the quantity $r^2 n^2(r)$ about the point $r = r_c$. The radicand then takes the form

$$r^2 n^2(r) - K^2 = r_c^2 n^2(r_c) - K^2 + \frac{d}{dr}\left[r^2 n^2(r)\right]_{r_c} (r - r_c)$$
$$+ \frac{1}{2}\frac{d^2}{dr^2}\left[r^2 n^2(r)\right]_{r_c} (r - r_c)^2 + O\left((r - r_c)^3\right). \qquad (11)$$

Simplifying (and neglecting extraneous multiplicative and additive constants) we find that, as indicated in Figure 3, if $\left(d\left[r^2 n^2(r)\right]/dr\right)_{r_c} > 0,$ then the integral in equation (9) has the functional form [16]

$$I \propto \int (r - r_c)^{-1/2} dr \propto (r - r_c)^{1/2} \to 0 \quad (12)$$

as $r \to r_c^+$. If on the other hand, $\left(d\left[r^2 n^2(r)\right]/dr\right)_{r_c} = 0,$ then

$$I \propto \int |r - r_c|^{-1} dr \propto \ln|r - r_c| \to -\infty \quad (13)$$

as $r \to r_c^+$.

(*Figures 3 & 4 near here*)

Generic $\eta(r)$ profiles for these two cases are illustrated schematically in Figures 3 and 4. In the monotonic case, the radius of closest approach for a given angle of incidence is denoted by $r_i$ in Figure 3; the distance of the ray trajectory from the center of the sphere is indicated on the *r*-axis. This is also indicated in the non-monotonic case in Figure 4. To interpret this Figure, it is best to consider rays with angles of incidence increasing away from zero. The radius (point) of closest approach increases in a continuous manner until $i = i_2$ as shown. At that stage the point of closest approach increases discontinuously by an amount $\Delta r$ to $r = r_c$, thereafter increasing continuously once again. This behavior corresponds to a spherical 'zone' of thickness $\Delta r$ into which *no rays* can penetrate. The situation is reversible: starting with $i = \pi/2$ and reducing it yields the same zonal gap.

In scattering theory, the logarithmic singularity (ii) above is associated with the phenomenon of *orbiting*. An extremum of $\eta(r)$ arises at $r = r_c$ when

$$n'(r_c) = -\frac{n(r_c)}{r_c} < 0, \quad (14)$$

meaning that the refractive index profile $n(r)$ either possesses a local minimum at $r = r_m > r_c$, or it tends monotonically to a constant value as *r* increases to one (see Figure 5). Of course, unlike the case of classical and/or atomic or molecular scattering, $n(r)$ and its corresponding potential $V(r)$ is in general piecewise continuous. The orbiting behavior illustrated in Figure 5 (lower figures) can be thought of as a type of 'mechanical' version of a limit cycle in a dynamical system. The connection between the two cases of 'classical' and 'potential' scattering is illustrated in Appendix 3.

(***Figure 5 near here***)

### 3. Analysis of specific profiles

We now examine two specific (and possibly singular) refractive index profiles for the *unit* sphere, generalizing somewhat that considered in [20]. Before so doing, we introduce some new notation. Electromagnetic waves possess two different polarizations: the transverse electric (*TE*) and transverse magnetic (*TM*) modes. Spherical *TE* modes have a magnetic field component in the direction of propagation, in this case that is in the radial direction, and spherical *TM* modes have an electric field component in the radial direction.

The first profile to be considered is

$$n(r) = n_1 r^{1/b-1} \left(2 - r^{2/b}\right)^{1/2}, \quad n_1 = n(1) > 1. \quad (15)$$

Note that if $b = 1$ and $n_1 = 1$, this profile corresponds to the classic Luneberg lens [21]. Using the result (5a) $D_0(i) = 2i - \pi + 2\Theta$, and substituting for $n(r)$ in the $\Theta$-integral, after some algebra the deviation angle can be shown to be

$$D_0(i) = \pi(b-1) + 2i - b\arcsin\left(\frac{\sin i}{n_1}\right). \quad (16)$$

For a zero-order bow to exist for some critical angle of incidence $i_c \in [0, \pi/2]$, it is necessary and sufficient that $D_0'(i_c) = 0$. This is the case if

$$\cos i_c = 2\left(\frac{n_1^2 - 1}{b^2 - 4}\right)^{1/2}, \quad (17)$$

which implies that $b \geq 2n_1$ if we restrict ourselves to the least potentially singular case of $b > 0$. We have therefore established that a zero bow can exist, unless $n_1 = 1$, whence equation (16) is a linear function of incidence angle $i$. It is interesting to note that the *TE* wave equation (see Appendix 2) has an exact solution for this choice of profile, finite for $0 \leq r \leq 1$, namely

$$S_l(r) = r^{l+1} \exp\left(-\frac{bkr^{2/b}}{2}\right) \times {}_1F_1\left(\frac{1}{2} + \frac{b}{2}\left(l + \frac{1}{2} - k\right); 1 + b\left(l + \frac{1}{2}\right); bkr^{2/b}\right). \quad (18)$$

Here $_1F_1$ refers to the confluent hypergeometric function. The *TM* equation cannot be expressed in terms of well-known functions, though it can be written in terms of generalized hypergeometric functions and solved by power series expansions in special cases. In a recent series of papers, Lock [4-6] analyzed the scattering of plane electromagnetic waves by a modified *Luneberg lens*. This 'lens' is a dielectric sphere of radius $a$ with a radially varying refractive index [21], specifically

$$n(r) = \frac{1}{f}\left[1 + f^2 - \left(\frac{r}{a}\right)^2\right]^{1/2}. \quad (19)$$

Here $f$ is a parameter determining the focal length of the lens. If $0 < f < 1$, the focus is inside the sphere (i.e. the focal length $< a$); for, $f = 1$ it is on the surface, and for $f > 1$ the focal point is outside the sphere. Note that, in contrast to the refractive index profiles (15) and (20), for the profile (19), $n(a) = 1$. Lock also found the existence of a transmission bow for this profile; indeed, this will occur for $f > 1$, whereas for $f = 1$ this bow evolves into an orbiting ray, and if $0 < f < 1$ this ray in turn evolves into a family of morphology-dependent resonances. In a wave theoretic approach to this problem [5], Lock studied the related radial 'Schrödinger' equation for the *TE* mode using the effective potential approach, discussed in Section 4.1 below.

When a family of rays has a near-grazing incidence on a dielectric sphere, the so-called 'far zone' consists of (i) an illuminated region containing rays refracted into the sphere and making $p-1$ internal reflections (where $p \geq 1$) before exiting the sphere, and (ii) a shadow zone into which no rays enter. (On a related topic, Lock showed that the asymptotic form of the Airy theory bow far into the illuminated region becomes the interference pattern of two supernumerary rays (with slightly different optical path lengths through the sphere). In an earlier paper [22] he showed that the zero ray/one ray transition for direct transmission is really a regular zero ray/two ray transition (as for a primary bow), with the second ray being a 'tunneling ray'; such tunneling will be discussed in section 4.1.)

The other choice for refractive index profile discussed here is

$$n(r) = \frac{2n_1 r^{1/c-1}}{1 + r^{2/c}}, \quad n_1 = n(1). \quad (20)$$

Detailed algebraic manipulation indicates that in this case,

$$D_0(i) = \pi(c-1) + 2i. \quad (21)$$

Obviously, $D'_0(i) \neq 0$ for any value of $i$, i.e. there is no zero-order bow for this profile! Both *TE* and *TM* modes have finite solutions for $0 \leq r \leq 1$, expressible in terms of the hypergeometric functions $_2F_1$, but we do not state them here. For the special case of $c = 1$ and $n_1 = 1$, this profile corresponds to the classic *Maxwell fish-eye* lens [23]. Other analytic solutions for the *TE/TM* modes will be discussed elsewhere [18].

## 4. Scattering by a transparent sphere: scalar wave description

The essential mathematical problem for scalar waves can be thought of either in terms of classical mathematical physics, e.g. the scattering of sound waves, or in quantum mechanical terms, e.g. the non-relativistic scattering of particles by a square potential well (or barrier) of radius $a$ and depth (or height) $V_0$ [7, 8]. In either case we can consider a scalar plane wave impinging in the direction $\theta = 0$ on a sphere of radius $a$. In what follows, a boldface letter refers to a vector quantity, thus here, $\mathbf{r} = \langle |r|, \theta, \phi \rangle$ (or $\langle r, \theta, \phi \rangle$) denotes a position vector in space (using a spherical coordinate system). Suppose that we had started with the 'classical wave equation' with dependent variable $\tilde{\psi}(\mathbf{r},t) = \psi(\mathbf{r})e^{-i\omega t}$. For the scalar electromagnetic problem, the angular frequency $\omega$, wavenumber $k$ and (constant) refractive index $n$ are related by $\omega = kc/n$, $c$ being the speed of light in vacuo. Then for a penetrable (= "transparent") sphere, the spatial part of the wave function $\psi(\mathbf{r})$ satisfies the scalar Helmholtz equation

$$\nabla^2 \psi + k^2 n^2 \psi = 0, \ r < a, \quad (22a)$$
$$\nabla^2 \psi + k^2 \psi = 0, \ r > a. \quad (22b)$$

Again, $k$ is the wavenumber and $n > 1$ is the (for now, constant) refractive index of the sphere. We can expand the wave function $\psi(\mathbf{r})$ as

$$\psi(\mathbf{r}) = \sum_{l=0}^{\infty} B_l(k) u_l(r) r^{-1} Y_l^m(\theta,\phi) \equiv \sum_{l=0}^{\infty} A_l(k) u_l(r) r^{-1} P_l(\cos\theta), \quad (23)$$

where $r = |\mathbf{r}|$, as noted above and the coefficients $A_l(k)$ will be 'unfolded' below (The coefficients $A_l$ and $B_l$ are related by a multiplicative normalization constant that need not concern us here.) The reason that the spherical harmonics $Y_l^m(\theta,\phi)$ reduce to the Legendre polynomials in the above expression is because the cylindrical symmetry imposed on the system by the incident radiation renders it axially symmetric (that is, independent of the azimuthal angle $\phi$). The equation satisfied by $u_l(r)$ is

$$\frac{d^2 u_l(r)}{dr^2} + \left[ k^2 - V(r) - \frac{l(l+1)}{r^2} \right] u_l(r) = 0, \quad (24)$$

where the potential $V(r)$ is now $k$-dependent, i.e.

$$V(r) = k^2(1 - n^2), \ r < a$$
$$V(r) = 0, \ r > a. \quad (25a, b)$$

Since $n > 1$ within the sphere, this potential corresponds to that of a spherical potential well of depth $V_0 = k^2(n^2 - 1)$. This leads very naturally to a discussion of the effective potential, wherein the potential $V(r)$ is combined with the 'centrifugal barrier' term $l(l+1)/r^2$.

**4.1 Morphology-dependent resonances: the effective potential $U_l(r)$ (constant $n$)**

A rather detailed study of the radial wave equations was carried out by Johnson [24], specifically for the *Mie solution* of electromagnetic theory (see Section 9). A crucial part of his analysis was the use of the effective potential for the *TE* mode of the Mie solution, but without any loss of generality we may still refer to the scalar problem here. This potential is defined as

$$U_l(r) = V(r) + \frac{l(l+1)}{r^2} = k^2(1-n^2) + \frac{l(l+1)}{r^2}, \quad r \leq a,$$
$$= \frac{l(l+1)}{r^2} \approx \frac{\lambda^2}{r^2}, \quad r > a. \tag{26a, b}$$

(*Figure 6 near here*)

It should be noted here that $\lambda$ as defined here is *not* the wavelength of the incident radiation. For large enough values of $l$, $[l(l+1)]^{1/2} \approx l + 1/2$. It is clear that $U_l(r)$ has a discontinuity at $r = a$ because of the 'addition' of a potential well to the centrifugal barrier. Thus there arises a tall and thin enhancement corresponding to a barrier surrounding a well (see Figure 6), and this suggests the possible existence of resonances, particularly between the top of the former and bottom of the latter, where there are three turning points (where the energy $k^2$ is equal to $U_l(r)$). Such resonances are called "shape resonances" (or sometimes "morphology-dependent resonances"); they are quasi-bound states in the potential well that escape by tunneling through the centrifugal barrier. The widths of these resonances depend on where they are located; the smaller the number of nodes of the radial wave function within the well, the deeper that state lies in the well. This in turn determines the width (and lifetime) of the state, because the tunneling amplitude is "exponentially sensitive" to the barrier height and width [13]. Since the latter decreases rapidly with the depth of the well, the smaller is the barrier transmissivity and the lowest-node resonances become very narrow for large values of $\beta = ka$. The lifetime of the resonance (determined by the rate of tunneling through the barrier) is inversely proportional to the width of the resonance, so these deep states have the longest lifetimes. (To avoid confusion of the node number $n$ with the refractive index in Figure 6, the latter has temporarily been written as *N*.)

Note that as $k^2$ is reduced, the bottom *B* of the potential rises (and for some value of $k$ the energy will coincide with the bottom of the well [24]); however, at the top of the well, $U_l(a) = \lambda^2/a^2$ is independent of $k^2$, but if $k^2$ is increased it will eventually coincide with the top of the well (*T*). Consider a value of $k^2$ between the top and the bottom of the well: within this range there will be three radial turning points, the middle one obviously occurring at $r = a$ and the largest at $r = b$ for which $U_l(b) = \lambda^2/b^2$. The smallest of the three ($r_{\min}$) is found by solving the equation

$$k^2 = \frac{\lambda^2}{r_{min}^2} - (n^2 - 1)k^2 \qquad (27)$$

to obtain, in terms of the impact parameter $b(\lambda) = \lambda / k,$

$$r_{min} = \frac{\lambda}{nk} \equiv \frac{b}{n}, \qquad (28)$$

By applying Snel's law for given *b*, it is readily shown that the distance of nearest approach of the equivalent ray to the center of the sphere is just $r_{min}$; indeed, there are in general many nearly-total internal reflections (because of internal incidence beyond the critical angle for total internal reflection) within the sphere between $r = b/n$ and $r = a.$ This is analogous to orbiting in a ray picture; on returning to its original location after one circumnavigation just below the sphere surface, a ray must do so with constructive interference. The very low leakage of these states allows the resonance amplitude and energy to build up significantly during a large resonance lifetime which in turn can lead to nonlinear optical effects. In acoustics these are called "whispering gallery modes".

The energy at the bottom of the well (i.e. $\lim_{r \to a^-} U_l(r)$) corresponding to the turning point at $r = a$ is determined by the impact parameter inequalities $a < b < na$, or in terms of $\lambda = kb$,

$$U_l(a^-) = \left(\frac{\lambda}{na}\right)^2 < k^2 < \left(\frac{\lambda}{a}\right)^2 = U_l(a^+). \qquad (29)$$

This is the energy range between the top and bottom of the well (and in which the resonances occur). To cross the "forbidden region" $a < r < b$ requires tunneling through the centrifugal barrier and near the resonance energies the usual oscillatory/exponential matching procedures can lead to very large ratios of internal to external amplitudes (see Figure 6(c)); these resonances correspond to "quasi-bound" states of electromagnetic radiation (that would be bound in the limit of zero leakage).

We now make a transition to discuss some of the related mathematical properties associated with resonances. In so doing, the reader should be alerted to a somewhat flexible notation used in connection with the scattering function (or *S*-matrix element to be discussed in Section 5), This is variously denoted by $\mathscr{S}_l(\lambda, k)$ or $\mathscr{S}_l(\beta)$, where $\beta = ka$, depending on the context. Mathematically, the resonances are complex eigenfrequencies associated with the poles $\lambda_n$ of the scattering function $\mathscr{S}_l(\lambda, k)$ in the first quadrant of the complex $\lambda$-plane; these are known as *Regge poles* (for real *k*). Corresponding to the energy interval $\left[U_l(a^-), U_l(a^+)\right]$, the real parts of these poles lie in the interval $(\beta, n\beta)$ (or equivalently, $(ka, nka)$); this corresponds to the tunneling region. The imaginary parts of the poles are directly related to resonance widths (and therefore lifetimes). As the node number *n* decreases, $\operatorname{Re}\lambda_n$ increases and $\operatorname{Im}\lambda_n$ decreases very rapidly (reflecting the exponential behavior of the barrier transmissivity). As $\beta$ increases, the poles $\lambda_n$ trace out Regge trajectories, and $\operatorname{Im}\lambda_n$ tend exponentially to zero. When $\operatorname{Re} \lambda_n$ passes close to a "physical" value, $\lambda = l + 1/2$, it is associated with a resonance in the $l^{th}$ partial wave; the larger the value of $\beta$, the sharper the resonance becomes for a given node number $n$.

## 5. Introduction to the scattering matrix.

The scattering matrix describes the relationship between the initial and final states of the 'system', whatever that may be. In fact it is very useful to relate these states at '$t = -\infty$' and '$t = \infty$' by means of the scattering operator $S$ acting on the wave function $\psi$, such that $\psi(\infty) = S\psi(-\infty)$. The matrix elements of the operator $S$ form the scattering matrix itself, not surprisingly.

Consider first, for simplicity, a scalar plane wave incident upon an *impenetrable* sphere of radius $a$. The solution of the Helmholtz equation (22b) (outside the sphere is) [7]

$$\psi_k(r,\theta) = \frac{1}{2}\sum_{l=0}^{\infty}(2l+1)i^l\left[h_l^{(2)}(kr) + \mathscr{S}_l(\beta)h_l^{(1)}(kr)\right]P_l(\cos\theta), \quad (30)$$

where $h_l^{(1)}(kr)$ and $h_l^{(2)}(kr)$ are spherical Hankel functions of the first and second kind respectively, and

$$\mathscr{S}_l(\beta) = -\frac{h_l^{(2)}(\beta)}{h_l^{(1)}(\beta)}; \quad \beta \equiv ka = \frac{2\pi a}{\lambda}. \quad (31)$$

The quantity $\mathscr{S}_l(\beta)$ is the element (for a given $l$-value) of the scattering or $S$-matrix. For 'elastic' (or non-absorptive) scattering, $\mathscr{S}_l(\beta)$ is a phase factor, and a very important one – it completely determines the nature of scattering in a potential field. As $|r| = r \to \infty$,

$$h_l^{(1)}(kr) \sim (-i)^{l+1}\frac{e^{ikr}}{kr}; \quad h_l^{(2)}(kr) \sim i^{l+1}\frac{e^{-ikr}}{kr}, \quad (32a, b)$$

hence inside the summation we have the term

$$\frac{(-1)^{l+1}}{kr}\mathscr{S}_l(\beta)\left[e^{ikr} + \frac{(-1)^{l+1}e^{-ikr}}{\mathscr{S}_l(\beta)}\right]. \quad (33)$$

Again, the reader should note that several possible contexts can be considered here. The modified partial wave number $\lambda = l + 1/2$ is in general considered to be complex, with $k$ being a real quantity, but here we consider $k$ to be a complex quantity also. Thus, so-called 'bound states' (of interest in quantum mechanics) are characterized by a pure imaginary wavenumber $k = ik_i$, $k_i > 0$ corresponding to energy $E = k^2 < 0$. In order for such a solution to be square-integrable in $(a, \infty)$, it is necessary that the second term vanish in equation (33) above. Formally, this will be the case if $\beta = ka$ is a pole of $\mathcal{S}_l(\beta)$. This is the essential significance of the poles of the *S*-matrix in what follows.

For a spherical square well or barrier, corresponding to a transparent sphere with constant refractive index *n*, the form of the scattering matrix elements for scalar waves is more complicated than (31). In fact [8; see also 25] in terms of spherical Bessel functions $(j_l)$ and spherical Hankel functions, the *S*-matrix is

$$\mathcal{S}_l(\beta) = -\frac{\beta j_l(\alpha) h_l^{'(2)}(\beta) - \alpha j'_l(\alpha) h_l^{(2)}(\beta)}{\beta j_l(\alpha) h_l^{'(1)}(\beta) - \alpha j'_l(\alpha) h_l^{(1)}(\beta)}. \quad (34)$$

Equation (34) is an expression of the matching at the finite boundary of the potential of the regular internal solution with the appropriate external solution of the Schrödinger equation. Using the notation of Nussenzveig [8], the expression (34) is equivalent to

$$\mathcal{S}_l(\beta) = -\frac{h_l^{(2)}(\beta)}{h_l^{(1)}(\beta)} \left[ \frac{\ln' h_l^{(2)}(\beta) - n \ln' j_l(\alpha)}{\ln' h_l^{(1)}(\beta) - n \ln' j_l(\alpha)} \right] \quad (35)$$

where ln' represents the logarithmic derivative operator, $j_l$ is a spherical Bessel function. The 'size parameter' $\beta = ka$ plays the role of a dimensionless external wavenumber, and $\alpha = n\beta$ is the corresponding *internal* wavenumber. Not surprisingly, $\mathcal{S}_l(\beta)$ may be equivalently expressed in terms of cylindrical Bessel and Hankel functions of half-integer order (see equation (39)). Note that for $l = 0$ the *S*-matrix element takes the simpler form [26]

$$\mathcal{S}_0(\beta) = e^{-2i\beta} \frac{\alpha \cot \alpha + i\beta}{\alpha \cot \alpha - i\beta}. \quad (36)$$

The $l^{th}$ "partial wave" in the series solution (23) (or (30)) is associated with an *impact parameter* $b(l) = (l + 1/2)/k$, i.e. only rays "hitting" the sphere $(b \leq a)$ are significantly scattered, and the number of terms that must be retained in the series to get an accurate result is slightly larger than *β*. Unfortunately, for visible light scattered by water droplets in the atmosphere, *β* is approximately several thousand and the partial wave series converges very slowly. This is certainly a non-trivial problem! In the next section we examine the resolution of this difficulty for both the scalar and the vector wave problem.

### 6. Introduction to complex angular momentum (CAM) theory: the Watson transform

In the early 20th Century there was a significant mathematical development that eventually had a profound impact on the study of scalar and vector scattering, and the present problem in particular. The *Watson transform*, originally introduced in 1918 by Watson in connection with the diffraction of radio waves around the earth, is a method for transforming the slowly-converging partial-wave series (e.g. (30)) into a rapidly convergent expression involving an integral in the complex angular-momentum plane. This allows the above transformation to effectively "redistribute" the contributions to the partial wave series into a few points in the complex plane – specifically the Regge poles and saddle-points. Such decomposition means that instead of identifying angular momentum with certain discrete real numbers, it is now permitted to vary continuously through complex values. However, despite this modification, the poles and saddle points have profound physical interpretations in the rainbow problem,

The Watson transform was subsequently modified by several mathematical physicists, including Nussenzveig [10, 12], in studies of the rainbow problem. It is intimately related to the *Poisson sum formula*

$$\sum_{l=0}^{\infty} g\left(l + \frac{1}{2}, x\right) = \sum_{m=-\infty}^{\infty} e^{-im\pi} \int_0^{\infty} g(\lambda, x) e^{2\pi im\lambda} d\lambda, \quad (37)$$

given an "interpolating function" $g(\lambda, x)$, where $x$ denotes a set of parameters and $\lambda = l + 1/2$ is again considered to be the complex angular momentum variable. The function $g$ is introduced to generate poles at the 'physical' values of $\lambda$ (or $l$) so that the corresponding residues account for the original partial wave series. By means of this conversion of a series to an integral in the complex plane, one is free to deform the path appropriately. The path can be chosen in such a way that the dominant high-frequency contributions to the radiation field come from a small number of 'critical points' (such as saddle points or complex poles). This avoids the complexity of summing these contributions over $\beta$ ($= ka$) partial waves (where $\beta \gg 1$).

It transpires that certain poles in the complex $\lambda$-plane are associated with surface waves (Regge poles; see below) and others are associated with morphology-dependent resonances in a particular partial wave. The latter are determined by the poles of the S-function in equation (34). But why is *angular momentum* the relevant parameter? A little physics helps us here. Although they possess zero rest mass, in terms of their associated de Broglie wavelength $\hat{\lambda}$, photons have energy $E = hc/\hat{\lambda}$ and momentum $E/c = h/\hat{\lambda}$, where $h$ is Planck's constant and $c$ is the speed of light in vacuo. (Note: the standard notation for wavelength is of course the Greek letter $\lambda$; here $\hat{\lambda}$ is used instead to avoid confusion with the complex angular momentum variable.) Thus for a non-zero impact parameter $b_i$ a photon will carry an angular momentum $b_i h / \hat{\lambda}$ ($b_i$ being the perpendicular distance of the incident ray from the axis of symmetry of the sun-raindrop system). Each of these discrete values can be identified with a term in the partial wave series expansion. Furthermore, as the photon undergoes repeated internal reflections, it can be thought of as orbiting the center of the raindrop. As will be re-emphasized below, the complex (Regge) poles mentioned above are associated with so-called 'creeping rays', generated by tangential incidence and propagating around the surface, shedding energy exponentially in a tangential direction. The damping is a result of the increasingly large imaginary part of these poles, leading to a rapidly convergent residue series in the shadow region (inhabited, not by real rays, but by diffracted rays). This approach works well for the impenetrable sphere discussed earlier. In the illuminated region the primary contributions come, not surprisingly, from real rays – stationary optical paths determined by Fermat's principle of least time. These rays are associated with stationary phase points on the real $\lambda$-axis (real saddle points).

Unfortunately, for a penetrable (or transparent, or dielectric) sphere, the Regge poles are situated much closer to the real $\lambda$-axis, and the convergence is compromised. To remedy this, the solution must be 'unfolded' in terms of surface-to-center reflections (and vice versa) – resulting in the so-called Debye series (see Appendix 1). The scattering amplitudes can then be expanded in a series, each term of which represents a surface interaction. When the modified Watson transform is applied to each term, one set of the resulting Regge-Debye poles, as they are called, are associated with rapidly damped surface waves see below), and rapidly convergent asymptotic expansions are obtained for each term in the Debye series. In this case, the critical points in the $\lambda$-plane are exactly those poles and (possibly complex) saddle points. There is a significant difference between the surface waves in this case and the case for the impenetrable sphere, however; they can also take a shortcut through the sphere (critical refraction) and re-emerge tangentially as surface waves.

For a Debye term of a given order $p$ (where $p-1$, $p \geq 1$) is the number of internal reflections at the surface, a primary rainbow (in particular) is associated in the $\lambda$-plane with the existence of two real saddle points that move towards each other as the 'rainbow scattering angle' is approached (see Figure 7), merging together at this angle, and beyond which (i.e. in the shadow region) the saddle points become complex and move away from the real axis in complex conjugate directions. Thus, as described in [7, 8, 10], from a mathematical point of view, *a rainbow can be defined as a collision between two saddle points in the complex angular momentum plane*.

As will be shown in Section 7, *the scattering amplitude* $f(k,\theta)$ is a quantity of fundamental importance in scattering theory, see Section 7 (see equations (50) and (51)). It is defined in terms of the scattering matrix elements $\mathscr{S}_l(k)$, and using the Poisson summation formula it may be recast as

$$f(k,\theta) = \frac{i}{ka} \sum_{m=-\infty}^{\infty} (-1)^m \int_0^{\infty} \left(1 - \mathscr{S}_l(\lambda,k)\right) P_{\lambda-1/2}(\cos\theta) e^{2\pi im\lambda} \lambda \, d\lambda. \quad (38)$$

*(Figure 7 near here)*

For fixed $\beta$, $\mathscr{S}_l(\lambda,\beta)$ is a meromorphic function of the complex variable $\lambda = l + 1/2$, and again it is the poles of this function that are of interest. In terms of cylindrical Bessel and Hankel functions, they are defined by the condition

$$\ln' H_\lambda^{(1)}(\beta) = n \ln' J_\lambda(\alpha). \quad (39)$$

As already noted, they are called Regge poles in the scattering theory literature [7, 8]. For the transparent sphere *two* types of Regge poles arise. Nussenzveig's *Class I poles* [9], located near the real $\lambda$-axis are associated with resonances, via the internal structure of the potential, which is now of course accessible. These are characterized by an effective radial wavenumber within the potential well. Typically, *Class II poles* are associated with surface waves for the impenetrable sphere problem mentioned above, – and lead to a rapidly convergent residue series, representing the surface wave (or diffracted or creeping ray) contributions to the scattering amplitude. Seeking poles of the *S*-matrix in the complex angular momentum plane, and their Regge trajectories as the energy *E* (or wavenumber *k*) is varied is in fact equivalent to analyzing these singularities and their trajectories in the complex *k*-plane as the angular momentum *l* is varied continuously through real values. In [25] it is pointed out that these two approaches – Regge trajectories and *k*-trajectories – are two different but complementary mathematical descriptions of the same physical phenomena, and that each one can provide insight into the other.

In the next section we examine another fundamental concept in scattering theory: the *phase shift*. This will prove to be crucial to understanding the changes induced on an incident wave on encountering a potential, be it of finite range or not.

## 7. The partial wave scattering phase shift $\delta_l(k)$

We return to the radial equation (24) in order to introduce this fundamental entity. The boundary conditions are that $u_l(r)$ and $u_l'(r)$ are continuous at the surface. We seek a solution satisfying the boundary condition at the origin

$$u_l(r)_{r \to 0} \sim r^{l+1}. \quad (40)$$

In the absence of a potential, the solutions $u_l(r)$ can be expressed in terms of Riccati-Bessel functions of the first and second kind (which are in turn related to the spherical Bessel functions of the first and second kind, $j_l(kr)$ and $y_l(kr)$ respectively):

$$\psi_l(kr) = kr j_l(kr) = \left(\frac{\pi kr}{2}\right)^{1/2} J_{l+1/2}(kr) \sim \sin(kr - l\pi/2) \text{ as } r \to \infty, \text{ and}$$

$$\xi_l(kr) = kr y_l(kr) = (-1)^{l-1} \left(\frac{\pi kr}{2}\right)^{1/2} Y_{(l+1/2)}(kr) \sim \cos(kr - l\pi/2) \text{ as } r \to \infty.$$
(41, 42)

(**Note:** some definitions of $\xi_l(kr)$ use the negative of the above expression, although $\chi_l(kr)$ is commonly used in the literature instead of $\xi_l(kr)$.) Based on the asymptotic forms of the Riccati-Bessel functions, we expect the solution of (24) to have the following property involving a $k$- and $l$-dependent phase shift:

$$u_l(r)_{r \to \infty} \sim \sin(kr - l\pi/2 + \delta_l(k)). \quad (43)$$

In fact, if $V(r)$ can be neglected for $r > r_0$, say, the solution of equation (24) can be written in terms of the phase shift $\delta_l(k)$ as [27, 28]

$$u_l(r) = kr\left[j_l(kr)\cos\delta_l(k) - y_l(kr)\sin\delta_l(k)\right]. \quad (44)$$

In particular, for a spherical well or barrier of radius $a$, the potential is zero for $r > a$. The $k$- or energy-dependent partial-wave phase shifts $\delta_l(k)$ represent the effect the potential $V(r)$ on the partial waves comprising the incident plane wave. The quantities $\delta_l(k)$ are real functions of the wave number $k$ when the potential $V(r)$, energy $E(=k^2)$ and angular momentum $l$ are all real. Shortly we shall reintroduce the S-matrix, this time with matrix elements defined in terms of the phase shifts $\delta_l(k)$. Particle scattering in a potential field is completely determined by these elements. The physical interpretation of the phase shifts can be understood as follows. The incoming plane wave is broken up into an infinite number of parts of differing angular momentum (these are the partial waves). Each partial wave interacts individually with the potential to produce a scattered outgoing partial wave. The phase of the outgoing wave is 'pushed out' by an amount delta by a repulsive potential, and the phase is 'pulled in' by an amount delta for an attractive potential. In optical terms for a sphere of refractive index $n > 1$, it is the latter case that applies: the potential is attractive.

Although it is the poles of the *S*-matrix that are of interest in this Chapter, it is valuable to reflect on the significance of several other concepts introduced here and below. As noted earlier, the phase shift is a measure of the departure of the radial wave function from the form it has when the potential $V(r)$ is zero. It follows from the definition below of the *K-matrix* that this too is a related measure of the distortion induced by a non-zero potential. The *K*-matrix is especially useful if the interaction is in some sense 'weak'. The *differential cross section* (equation (52b)) is useful because it is the quantity that is directly measured in scattering experiments. The *Jost functions* are useful because they help express the pole structure and associated zero structure of the *S*-matrix in a very straightforward way.

Returning to the asymptotic result (43), it is also of interest to note that it can be expressed in two other equivalent ways. They are

$$(i) \ u_l(r)_{r \to \infty} \sim \cos \delta_l \left[ \sin(kr - l\pi/2) + K_l \cos(kr - l\pi/2) \right],$$

$$\text{and } (ii) \ u_l(r)_{r \to \infty} \sim \frac{e^{-i\delta_l}}{2i} \left[ e^{-i(kr - l\pi/2)} - \mathscr{S}_l(k) e^{i(kr - l\pi/2)} \right]. \quad (45, 46)$$

The first of these equations defines the elements of the *K*-matrix, i.e. $K_l = \tan \delta_l$, and the second (re)defines the *S*-matrix elements, i.e. $\mathscr{S}_l(k) = e^{2i\delta_l}$. In fact,

$$\mathscr{S}_l(k) = \exp\left[2i\delta_l(k)\right] = \frac{1 + i \tan \delta_l(k)}{1 - i \tan \delta_l(k)} \equiv \frac{1 + iK_l(k)}{1 - iK_l(k)}. \quad (47)$$

The integral equation satisfied by the radial wave function $u_l(r)$ can also be written in terms of the Riccati-Bessel functions as follows:

$$u_l(r) = \psi_l(kr) - k^{-1} \int_0^r \left[ \psi_l(kr)\xi_l(kr') - \psi_l(kr')\xi_l(kr) \right] V(r')u_l(r')dr'. \quad (48)$$

This may be verified by direct substitution into equation (24), where now

$$\lim_{r \to 0} u_l(r) = \lim_{r \to 0} \psi_l(kr) \to \frac{(kr)^{l+1}}{(2l+1)!!}. \quad (49)$$

At large distances from the sphere $(r \gg a)$ the complete wave field $\psi(r)$ can be decomposed into an (axially symmetric) incident wave + scattered field, i.e.

$$\psi(r,\theta) \sim e^{ikr\cos\theta} + \frac{f(k,\theta)}{r} e^{ikr}. \quad (50)$$

In terms of the scattering matrix element for a given *l*, and therefore $\mathscr{S}_l(k)$, the *scattering amplitude* is defined as

$$f(k,\theta) = (2ik)^{-1} \sum_{l=0}^{\infty} (2l+1)(\mathscr{S}_l(k)-1) P_l(\cos\theta). \quad (51)$$

$P_l(\cos\theta)$ is a Legendre polynomial of degree *l*. In terms of the phase shift $\delta_l$, the scattering amplitude can be written as

$$f(k,\theta) = k^{-1} \sum_{l=0}^{\infty} (2l+1) e^{i\delta_l} \sin\delta_l P_l(\cos\theta); \quad (52a)$$

For completeness, in the scattering literature, the *differential scattering cross section* is defined by

$$\frac{d\sigma}{d\Omega} = \frac{\text{scattered flux/unit solid angle}}{\text{incident flux/unit area}} = |f(\theta)|^2, \quad (52b)$$

and the *total* (*elastic*) *cross section* $\sigma$ is obtained by integrating the differential cross section over all scattering angles, i.e.

$$\sigma = \int_0^{2\pi} d\phi \int_0^{\pi} |f(\theta)|^2 \sin\theta d\theta = 2\pi \int_0^{\pi} |f(\theta)|^2 \sin\theta d\theta. \quad (52c)$$

The quantity

$$p_l(k^2) = k^{-1} e^{i\delta_l} \sin\delta_l = (2ik)^{-1} (e^{2i\delta_l} - 1) \quad (53)$$

is often referred to as the *partial wave scattering amplitude*.

## 8. Analytic properties of the *S*-matrix: the Jost functions

We now consider in more detail the analytic properties of the partial wave *S*-matrix, with elements defined by equations (34) or (35) (for example), in the complex momentum plane. We can show that the poles of the *S*-matrix lying on the positive imaginary *k*-axis correspond to bound states, while poles lying in the lower half *k*-plane close to the positive real *k*-axis correspond to the resonances discussed above (see Appendix 4). We may also derive an expression for the behavior of the phase shift and the cross section when the energy of the scattered particle is in the neighborhood of these poles. Consider again the solution $u_l(r)$ of the radial Schrödinger equation (24) describing the scattering of a particle by a spherically symmetric potential $V(r)$. Implicit in the results to be stated here are certain requirements on the potential $V(r)$. It must be a real, almost everywhere continuous function vanishing at infinity. Furthermore [29, 30], it must be the case that

$$(i) \int_c^\infty |V(r)| \, dr = M(c) < \infty \text{ and}$$

$$(ii) \int_0^{c'} r|V(r)| \, dr = N(c') < \infty,$$

where $c$ and $c'$ are positive constants (but otherwise arbitrary). The first of these conditions is equivalent to $V \sim r^{-(1+\varepsilon)}$ as $r \to \infty$, $\varepsilon > 0$ (i.e. $rV(r) \to 0$ as $r \to \infty$), and the second implies that $V \sim r^{-(2+\varepsilon')}$ as $r \to 0$, $\varepsilon' > 0$ (i.e. $r^2 V(r) \to 0$ as $r \to 0$). (Note that Burke [27a] places more stringent conditions on the potential for the existence of bound states; instead of (i) he requires that $\int_0^\infty r^2 |V(r)| \, dr < \infty$). We also introduce two (normalized) *Jost solutions* $f_l(\pm k, r)$ of (24), defined by the relations

$$\lim_{r \to \infty} f_l(\pm k, r) e^{\pm i(kr \mp l\pi/2)} = 1. \quad (54)$$

This condition at infinity defines $f_l(k, r)$ uniquely in the lower half $k$-plane, where it is analytic. In the upper half-plane $f_l(k, r)$ is no longer unique because it is always possible to add to it a term proportional to the other Jost solution $f_l(-k, r)$. If the potential vanishes identically beyond a certain distance $a$ then the $f_l(\pm k, r)$ are analytic functions of $k$ in the open $k$-plane for all fixed values of $r$, that is, they are entire functions of $k$. We can express the physical solution of (24), defined by the boundary conditions as a linear combination of $f_l(\pm k, r)$, in keeping with the form (44). Thus

$$u_l(r) \propto \left[ f_l(k, r) + (-1)^{l+1} f_l(-k, r) \mathscr{S}_l(k) \right]. \quad (55)$$

From a theorem proved by Poincaré, the absence of a $k$-dependence in this boundary condition implies that this solution is an entire function of $k$. The *Jost functions* are then defined by

$$\tilde{f}_l(\pm k) = W\left[ f_l(\pm k, r), u_l(r) \right], \quad (56)$$

where the Wronskian $W$ is independent of $r$. It is also convenient to introduce a *normalized* Jost function $f_l(\pm k)$ by

$$f_l(\pm k) = \frac{k^l \exp(\pm i l\pi / 2)}{(2l+1)!!} \tilde{f}_l(\pm k). \quad (57)$$

(Note that the notation for these functions should not be confused with the definition of the scattering amplitude in equations (51) and (52a)). The functions $f_l(+k)$ and $f_l(-k)$ are continuous at $k = 0$ and approach unity at large $|k|$ for $\text{Im } k \leq 0$ and $\text{Im } k \geq 0$, respectively.

Since

$$W[f_l(\pm k, r), f_l(\mp k, r)] = \pm 2ik, \quad (58)$$

$u_l(r)$ may be written in the form

$$u_l(r) = \frac{1}{2ik}[\tilde{f}_l(k)f_l(-k,r) - \tilde{f}_l(-k)f_l(k,r)]. \quad (59)$$

Comparing this equation with the asymptotic form (44) and using (54) then yields the following expression for the S-matrix elements:

$$\mathscr{S}_l(k) = e^{i\pi l}\frac{\tilde{f}_l(k)}{\tilde{f}_l(-k)} = \frac{f_l(k)}{f_l(-k)}. \quad (60)$$

This equation relates the analytic properties of the S-matrix with the simpler analytic properties of the Jost functions [27b]. Since, in particular, $f_l(-k, r)$ satisfies equation (24), i.e.

$$\left(\frac{d^2}{dr^2} + k^2 - V(r) - \frac{l(l+1)}{r^2}\right)f_l(-k, r) = 0. \quad (61)$$

It follows that if we now take the complex conjugate of this equation, we obtain (for real $l$ and $V(r)$)

$$\left(\frac{d^2}{dr^2} + \overline{k}^2 - V(r) - \frac{l(l+1)}{r^2}\right)\overline{f}_l(-k, r) = 0. \quad (62)$$

If we also let $k \to -\overline{k}$ in (61) we also have that

$$\left(\frac{d^2}{dr^2} + \overline{k}^2 - V(r) - \frac{l(l+1)}{r^2}\right)f_l(\overline{k}, r) = 0. \quad (63)$$

Furthermore,

$$\overline{f}_l(-k, r)_{r\to\infty} \sim \exp(-i\overline{k}r) \text{ and } f_l(\overline{k}, r)_{r\to\infty} \sim \exp(-i\overline{k}r), \quad (64a, b)$$

i.e. they satisfy the same boundary conditions at infinity. Since these functions also satisfy the same differential equation, namely (62) and (63), respectively, they are equal for all $r$ for all points in the upper half $k$-plane and for all other points which admit an analytic continuation from the upper half $k$-plane.

Hence in this region $\overline{f_l}(-k,r) = f_l(\overline{k},r)$, and hence, from (56), $\overline{\tilde{f}_l}(-k) = \tilde{f}_l(\overline{k})$. Therefore from (60) we find that

$$\mathscr{S}_l(k)\mathscr{S}_l(-k) = e^{2\pi i l} \frac{\tilde{f}_l(k)}{\tilde{f}_l(-k)} \frac{\tilde{f}_l(-k)}{\tilde{f}_l(k)} = 1. \quad (65)$$

We also have the unitarity condition

$$\mathscr{S}_l(k)\overline{\mathscr{S}_l(\overline{k})} = \frac{\tilde{f}_l(k)}{\tilde{f}_l(-k)} \frac{\overline{\tilde{f}_l(\overline{k})}}{\overline{\tilde{f}_l(-k)}} = 1. \quad (66)$$

These relations give in turn the reflection property

$$\mathscr{S}_l(k) = e^{2\pi i l} \overline{\mathscr{S}_l(-\overline{k})}. \quad (67)$$

It follows from (66) that if $k$ is real then $|\mathscr{S}_l(k)| = 1$ and in terms of the real phase shift $\delta_l(k)$,

$$\mathscr{S}_l(k) = \exp[2i\delta_l(k)]. \quad (68)$$

This is a result already noted above. The poles and zeros of the *S*-matrix are symmetrically situated with respect to the imaginary *k*-axis, because it follows from (67) that if the *S*-matrix has a pole at the point $k$, then it also has a pole at the point $-\overline{k}$ and from (65) and (66) it has zeros at the points $-k$ and $\overline{k}$. For potentials satisfying the conditions stated at the beginning of this section, only a finite number of bound states can be supported and these give rise to the poles lying on the positive imaginary axis in Figure 8. However, an infinite number of poles can occur in the lower half *k*-plane. If they do not lie on the negative imaginary *k*-axis, they occur in pairs symmetric with respect to this axis, as discussed above. If they lie on the negative imaginary *k*-axis, they are often referred to as *virtual state* poles; the wave functions corresponding to these states cannot be normalized. Poles lying in the lower half *k*-plane and close to the real positive *k*-axis give rise to resonance effects in the cross section equation (52c). Poles lying in the lower half *k*-plane and far away from the real positive *k*-axis contribute to the smooth "background" or "non-resonant" scattering. The distribution of poles in the complex *k*-plane has been discussed in detail in a few cases, (see e.g. [26]) for scattering by a square well potential.

(*Figure 8 near here*)

### 8.1 The Breit-Wigner form

Consider an isolated pole in the *S*-matrix which lies in the lower half *k*-plane close to the positive real *k*-axis. This pole gives rise to resonance scattering at the nearby real energy. We note (by virtue of Appendix 5) that the pole occurs at the complex energy

$$E = E_r - \frac{i}{2}\Gamma, \quad (69)$$

where $E_r$ is the resonance position, and $\Gamma$ is the resonance width, and both are real positive numbers. From the unitarity relation (66) we see that corresponding to this pole there is a *zero* in the S-matrix (at a complex energy given by $E = E_r + i\Gamma/2$) in the upper half k-plane. For energies $E$ on the real axis in the neighborhood of this pole, the S-matrix can be written in a form which is both unitary and explicitly contains the pole and zero:

$$\mathscr{S}_l(k) = \exp\left[2i\delta_l^0(k)\right]\frac{E - E_r - i\Gamma/2}{E - E_r + i\Gamma/2}. \quad (70)$$

The quantity $\delta_l^0(k)$ in this equation is called the "background" or "non-resonant" phase shift. Provided that the energy $E_r$ is not close to threshold, $E = 0$, nor to another resonance then the background phase shift is slowly varying with energy. Comparing (68) and (70) we obtain the following expression for the phase shift:

$$\delta_l(k) = \delta_l^0(k) + \delta_l^r(k). \quad (71)$$

The quantity

$$\delta_l^r(k) = \arctan\left(\frac{\Gamma/2}{E_r - E}\right) \quad (72)$$

is called the "resonant" phase shift which is seen to increase through $\pi$ radians as the energy $E$ increases from well below to well above the resonance position $E_r$.

### 8.2 Further comments on Jost functions and bound states

It can be seen from equation (46) that $\mathscr{S}_l(\beta)$ is proportional to the ratio of the coefficients of the outgoing and incoming waves (recall that the harmonic factor $e^{-i\omega t}$ has been suppressed). According to the theorem of Poincaré mentioned earlier, if the boundary conditions on a differential equation are independent of the parameters in the equation, the solutions will be analytic functions of those parameters. Therefore, the solutions $u_l(r)$ of equation (24) will be analytic functions of energy $E = k^2$ if the normalization condition on the behavior of $r^{-(l+1)}u_l(r)$ as $r \to 0$ is also independent of $k^2$ [31, 32]. For small values of $k$ it can be shown that $\tan\delta_l \sim k^{2l+1} = (k^2)^{l+1/2}$; in other words, $\delta_l$ is an analytic function of $k$ (as opposed to $k^2$) near zero energy. Since $\exp(2i\delta_l)$ is an analytic function of $\delta_l$, the Jost functions will share the branch points of $\delta_l$. As noted earlier, it is customary to divide the $k^2$-plane into two Riemann sheets by requiring that the 'physical' sheet corresponds to $\operatorname{Im} k = \operatorname{Im}(k^2)^{1/2} > 0$ on that sheet. The positive $k^2$-axis is a branch cut [33].

From equation (60), poles of $\mathscr{S}_l(k)$ occur when $f_l(-k) = 0$. In the neighborhood of such a zero, we see from equations (54) and (59) that, asymptotically

$$u_l(r) \propto f_l(k) e^{ikr}. \quad (73)$$

Recalling that on the physical sheet $\operatorname{Im} k = k_i > 0$ it follows that $u_l(r) \propto e^{ik_r r} e^{-k_i r}$ so that it is a square integrable and hence normalizable solution; this means it represents a bound state. But such a state for an attractive potential (such as a spherical square well) implies that $k^2 < 0$, that is, $k = ik_i$. Poles on the physical sheet produce an exponentially decaying wave function, so the zeros of $f_l(-k)$ for $k_i > 0$ are bound states. In particular, for the case $l = 0$ it can be shown that $\mathscr{S}_l(k)$ can have poles only where either $\operatorname{Re}(k) = 0$ or $\operatorname{Im}(k) < 0$ [26, 34, 43] (this is proved in Appendix 4). Furthermore, since the partial wave amplitude $p_l(k^2)$ can be expressed in terms of the Jost functions, this means that poles of $p_l(k^2)$ (equation (53)) on the physical sheet are also associated with bound states.

In summary at this point, the scattering matrix elements $\mathscr{S}_l(k)$, regarded as functions of the complex variable $k$, have several valuable physical interpretations. If $k$ is real, the scattering is defined in terms of real phase shifts $\delta_l$ which in turn determine the scattering cross section. Poles of the elements which are pure imaginary with (i) $\operatorname{Im}(k) > 0$ correspond to bound states of the potential, those with (ii) $\operatorname{Im}(k) < 0$ correspond to 'virtual' or non-normalizable states (or 'antibound' states [26]). If the poles are complex with $\operatorname{Im}(k) < 0$ they are sometimes referred to as 'quasi-stationary states', and if $\operatorname{Re}(k) > 0$ and $|\operatorname{Im}(k)| \ll 1$ they are called resonance poles. In the complex $E$-plane, poles associated with quasi-stationary states are on the second sheet of the Riemann energy surface.

### 8.3 Regge poles and Regge trajectories

Following directly from the previous sentence, the 'unphysical' Riemann sheet (but close to the branch cut), poles of the $S$-matrix elements (now written as $\mathscr{S}_l(k^2)$) i.e. at $E = k^2 = k_r^2 - i\Gamma/2$ (where $\Gamma$ is 'small' and positive) give rise to the familiar Breit-Wigner expression examined above for the phase shift $\delta_l$. Each such pole on the unphysical sheet corresponds to a *resonance* with energy $k_r^2$ and 'half-width' $\Gamma/2$. What happens as $l$ varies in the radial Schrödinger equation? Again, from Poincaré's theorem, the Jost functions will be analytic functions of $l$ as well as $k^2$, and we know that the bound states of $V(r)$ are found as the zeros of $f_l(-k)$. This criterion can be regarded as an implicit function in $l$ and $k$ (or indeed, $l$ and $E = k^2$), i.e. $l = g(E)$ (this is a generic function, not the same one as in equation (37)). Again, equations (54) and (59) imply that for $k^2 < 0$

$$u_l(r)_{r\to\infty} \propto \left(f_l(k)e^{-k_i r} - f_l(-k)e^{k_i r}\right). \quad (74)$$

Since the radial Schrödinger equation is expressed in terms of real quantities only so the solution $u_l(r)$ is real and so are the Jost functions by virtue of (74); therefore (in particular) $f_l(-k)$ is also real. Hence the zeros $l = g(E)$ of this function must also be real functions. On the other hand, if $k^2 > 0$, $u_l(r)$ is still real, but the complex exponential factors imply that $f_l(-k)$ will be a complex function, whence in general, the Regge pole trajectories $l = G(E)$ (say) will be complex. However, bound states of angular momentum $l$ exist when a trajectory intersects the line $l = m$, $m = 0, 1, 2, ...$, with corresponding energy $k^2 = k^2(l)$.

By contrast, in the complex $k$-plane for real and positive values of $\lambda = l + 1/2$, all poles in the upper half-plane must lie on the imaginary axis. Both complex and pure imaginary poles can be present in the lower half-plane [26], and for physical (half-integer) values of $\lambda$, the symmetry of these poles with respect to the imaginary axis is established from the following property for the generalized $S$-matrix element $\mathscr{S}(\lambda, k)$, namely that $\mathscr{S}(\lambda, k) = \overline{\mathscr{S}(\overline{\lambda}, -\overline{k})}$. Note that, according to [25] this relation is no longer valid for unphysical values of $\lambda$. In summary, there are two infinite families of '$k$-poles' (corresponding to the two classes of Regge poles discussed by Nussenzveig [9, 10]; see also Section 6 above). Class I poles, we recall, are determined by the interior of the potential, and are located in the fourth quadrant near the positive real semi-axis. By contrast, class II poles correspond to surface modes on the 'spherical potential', and are located in the third and fourth quadrants. More details can be found in [25].

## 9. The vector problem: the Mie solution of electromagnetic scattering theory

The quantum mechanical scalar analysis in previous sections is appropriate primarily for non-relativistic scattering of a projectile 'particle' of mass *m*. In this section a very different phenomenon is discussed: scattering of zero rest-mass photons. The crucial point to note here is that both of these very different physical systems share the same mathematical structure, namely the properties of the scalar wave equation.

So having made considerable reference to the scalar problem and its connection with the potential scattering theory, we now turn to the vector problem which for electromagnetic waves possesses two polarizations (the *TE* and *TM* modes); each radial equation can be examined in turn as a scalar problem. Mie theory is based on the solution of Maxwell's equations of electromagnetic theory for a monochromatic plane wave from infinity incident upon a homogeneous isotropic sphere of radius *a*. The surrounding medium is transparent (as the sphere may be), homogeneous and isotropic. The incident wave induces forced oscillations of both free and bound charges in synchrony with the applied field, and this induces a secondary electric and magnetic field, each of which has components inside and outside the sphere [35].

In this Section reference will be made to the intensity functions $i_1$, $i_2$, the Mie coefficients $a_l, b_l$ and the angular functions $\pi_l, \tau_l$. The intensity functions are proportional to the square of the magnitude of two incoherent, plane-polarized components scattered by a single particle; they are related to the scattering amplitudes $S_1$ and $S_2$ in the notation of Nussenzveig [11]. The function $i_1(\beta, n, \theta)$ is associated with the electric oscillations perpendicular to the plane of scattering (sometimes called horizontally polarized) and $i_2(\beta, n, \theta)$ is associated with the electric oscillations parallel to the plane of scattering (vertically polarized). The scattered spherical wave is composed of an infinite number of partial waves, the amplitudes of which depend on $a_l(\beta, n)$ and $b_l(\beta, n)$. In physical terms, these may be interpreted as the $l^{th}$ electrical and magnetic multipole waves respectively. The first set is that part of the solution for which the radial component of the magnetic vector in the incident wave is zero; in the second set the corresponding radial component of the electric vector is zero. A given partial wave can be thought of as coming from an electric or a magnetic multipole field, the first wave coming from a dipole field, the second from a quadrupole, and so on [35]. The angular functions $\pi_l(\cos\theta)$ and $\tau_l(\cos\theta)$ are, as their name implies, independent of size ($\beta$) and refractive index ($n$).

For a point $P$ located a distance $r$ from the origin of coordinates, at polar angle $\theta$ and azimuthal angle $\varphi$ the scattered intensities $I_\theta$ and $I_\phi$ are respectively

$$I_\theta = i_2\left(\frac{1}{kr}\right)^2 \cos^2\phi \text{ and } I_\phi = i_1\left(\frac{1}{kr}\right)^2 \sin^2\phi, \quad (75a, b)$$

where $i_j = |S_j|^2$, $j = 1, 2$ and the amplitude functions $S_j$ are given by

$$S_1 = \sum_{l=1}^{\infty} \frac{2l+1}{l(l+1)} \left[a_l \pi_l(\cos\theta) + b_l \tau_l(\cos\theta)\right], \text{ and}$$

$$S_2 = \sum_{l=1}^{\infty} \frac{2l+1}{l(l+1)} \left[a_l \tau_l(\cos\theta) + b_l \pi_l(\cos\theta)\right].$$

(76a, b)

$l$ is the order of the induced electric or magnetic multipole. The Mie angular functions $\pi_l(\cos\theta)$ and $\tau_l(\cos\theta)$ are defined in terms of the associated Legendre functions of the first kind, $P_l^1(\cos\theta)$ as

$$\pi_l(\cos\theta) = \frac{P_l^1(\cos\theta)}{\sin\theta} \text{ and } \tau_l(\cos\theta) = \frac{d}{d\theta} P_l^1(\cos\theta). \quad (77a, b)$$

The scattering coefficients $a_l$ and $b_l$ are defined in terms of the previously encountered Riccati-Bessel functions of the first and second kinds respectively. $a_l$ and $b_l$ can be written in terms of the Riccati-Hankel function of the first kind, $\zeta_l^{(1)}(z) = z h_l^{(1)}(z) = \psi_l(z) + i\xi_l(z)$, i.e.

$$a_l = \frac{\psi_l(\beta)\psi_l'(\alpha) - n\psi_l(\alpha)\psi_l'(\beta)}{\zeta_l^{(1)}(\beta)\psi_l'(\alpha) - n\psi_l(\alpha)\zeta_l^{(1)'}(\beta)} \quad \text{and}$$

$$b_l = \frac{\psi_l(\alpha)\psi_l'(\beta) - n\psi_l(\beta)\psi_l'(\alpha)}{\zeta_l^{(1)'}(\beta)\psi_l(\alpha) - n\psi_l'(\alpha)\zeta_l^{(1)}(\beta)}.$$

(78a, b)

For future reference, the Riccati-Hankel function of the second kind is defined by $\zeta_l^{(2)}(z) = zh_l^{(2)}(z) = \psi_l(z) - i\xi_l(z)$. The dimensionless size parameters $\beta = ka$ and $\alpha = n\beta$ are again used in equations (78a, b). These expressions can be simplified by the introduction of phase shift angles; and results in considerable simplification if the refractive index is real [36]. In [36] it is demonstrated that the Mie formulae lead, for large values of $\beta$, to a principle for localizing rays and separating diffracted, refracted and reflected light (in the sense of geometrical optics). The principle asserts that the term of order $l$ in the partial wave expansion corresponds approximately to a ray of distance $(l+1/2)/k$ from the center of the particle (this is just the impact parameter). When $\beta \gg 1$, the expansions for the $S_j$ $(j = 1, 2)$ may be truncated at $l + 1/2 \approx \beta$ (in practice, $l_{max} \sim \beta + 4\beta^{1/3} + 2$; see [8, 9, 37]), and the remaining sum is separated into two parts: a diffracted light field component independent of the nature of the particle, and reflected and refracted rays dependent on the particle (see also [38]).

From (78a, b) above we can define the new quantities [7]

$$P_l^e \equiv \psi_l(\beta)\psi_l'(\alpha) - n\psi_l(\alpha)\psi_l'(\beta),$$
$$Q_l^e \equiv \xi_l(\beta)\psi_l'(\alpha) - n\psi_l(\alpha)\xi_l'(\beta),$$
$$P_l^m \equiv \psi_l(\alpha)\psi_l'(\beta) - n\psi_l(\beta)\psi_l'(\alpha),$$
$$Q_l^m \equiv \xi_l'(\beta)\psi_l(\alpha) - n\psi_l'(\alpha)\xi_l(\beta).$$

(79a – d)

The notation of Grandy [7] is followed here (but a common alternative notation is *N/D* rather than *P/Q*). These quantities are real if $n$ is real. Then the external coefficients (in particular) may be written as

$$a_l = \frac{P_l^e}{P_l^e + iQ_l^e}, \quad b_l = \frac{P_l^m}{P_l^m + iQ_l^m}. \quad (80a, b)$$

Furthermore, we may define (for real $n$) the real phase shifts $\delta_l$ in terms of the *K*-matrix elements

$$\tan \delta_l^e \equiv \frac{P_l^e}{Q_l^e} \quad \text{and} \quad \tan \delta_l^m \equiv \frac{P_l^m}{Q_l^m}. \quad (81a, b)$$

Hence

$$a_l = \frac{1}{2}\left[1 - \exp(2i\delta_l^e)\right], \quad b_l = \frac{1}{2}\left[1 - \exp(2i\delta_l^m)\right]. \quad (82a, b)$$

Also it is readily shown that

$$a_l = \frac{(P_l^e)^2}{(P_l^e)^2 + (Q_l^e)^2} - i\frac{P_l^e Q_l^e}{(P_l^e)^2 + (Q_l^e)^2}, \quad (83)$$

from which it follows that, for no absorption (i.e. elastic scattering)

$$\text{Re}(a_l) = |a_l|^2 = \sin^2 \delta_l^e \in [0,1], \text{ and } \text{Im}(a_l) = \frac{1}{2}\sin 2\delta_l^e \in \left[-\frac{1}{2}, \frac{1}{2}\right]. \quad (84\text{a, b})$$

A similar set of equations can be deduced for $b_l$. It is interesting to note that the locus of $a_l$ and $b_l$ in the complex $\delta_l$-plane is a circle of radius ½ with center at $(1/2, 0)$. The scalar *partial* scattering amplitudes $f_l(k)$ can be defined using equation (51) as

$$f_l(k) = \frac{e^{-il\pi/2}}{2ik}\left[\mathcal{S}_l(k) - 1\right], \quad (85)$$

(on reverting to the former notation for $\mathcal{S}_l(k)$), where $\mathcal{S}_l(k) = \exp(2i\delta_l)$, the vector problem can be characterized by (for real $n$) the unitary matrix

$$\mathcal{S}_l = \begin{pmatrix} \mathcal{S}_l^e & 0 \\ 0 & \mathcal{S}_l^m \end{pmatrix}. \quad (86)$$

If we now write

$$a_l = \frac{1}{2}\left[1 - \mathcal{S}_l^e(k)\right], \quad b_l = \frac{1}{2}\left[1 - \mathcal{S}_l^m(k)\right], \quad (87\text{a, b})$$

Substitution into (82a, b) yields the expressions in terms of $\alpha$ and $\beta$

$$\mathcal{S}_l^e(k) = -\frac{\zeta_l^{(2)}(\beta)}{\zeta_l^{(1)}(\beta)}\left[\frac{\ln'\zeta_l^{(2)}(\beta) - n^{-1}\ln'\psi_l(\alpha)}{\ln'\zeta_l^{(1)}(\beta) - n^{-1}\ln'\psi_l(\alpha)}\right],$$

$$\mathcal{S}_l^m(k) = -\frac{\zeta_l^{(2)}(\beta)}{\zeta_l^{(1)}(\beta)}\left[\frac{\ln'\zeta_l^{(2)}(\beta) - n\ln'\psi_l(\alpha)}{\ln'\zeta_l^{(1)}(\beta) - n\ln'\psi_l(\alpha)}\right]. \quad (88\text{a, b})$$

In these expressions the notation $\ln' f(z) = d(\ln f(z))/dz$ has been used. As we have seen, $\text{Re}(a_l)$ reaches its maximum value (unity) when $Q_l^e = 0$ (for the *TM* modes), and similarly, a maximum occurs for $\text{Re}(b_l)$ when $Q_l^m = 0$ (*TE* modes). These conditions correspond to Johnson's condition for resonance [24], and as Grandy [7] shows in some detail, they are also equivalent to the poles of the Mie coefficients $a_l$ and $b_l$ in the complex $\beta$-plane, which are *in turn* equivalent to the poles of the scattering matrix elements $\mathscr{S}_l^m(\lambda,\beta)$ and $\mathscr{S}_l^e(\lambda,\beta)$ in the complex $\lambda$-plane. A valuable examination of the formal analogies between Mie theory and time-independent quantum scattering by a radial potential for both transparent and absorbing 'particles' has been carried out in [39].

Solutions of the radial (Debye) equation (24) are linear combinations of the Riccati-Bessel functions $\psi_l(kr)$ and $\xi_l(kr)$ which vanish at the origin and match appropriately at $r = a$, i.e.

$$u_l^v(r) \propto \psi_l(nkr), \ 0 \leq r \leq a, \text{ and}$$

$$u_l^v(r) \propto \left( \xi_l(kr) - \frac{Q_l^v}{P_l^v} \psi_l(kr) \right), \ r \geq a. \quad (89\text{a, b})$$

The superscript $v = e$ or $m$ refers respectively to the electric or magnetic multipole modes respectively. Within the barrier, the solution $u_l^v(r)$ must be exponentially increasing with $r$, from which we infer that $Q_l^e = 0$ for the *TM* modes and $Q_l^m = 0$ for the *TE* modes. As pointed out in [7], these conditions determining the discrete 'energy levels' of a resonance are precisely the conditions mentioned above.

**Conclusion**

This article attempts to categorize and summarize some of the many and various connections that exist between ray theory, wave theory and potential scattering theory. By 'meandering' through these related areas in the broader field of mathematical physics, it is hoped that the reader will recognize how each of the levels of description can inform the others, resulting (it is to be hoped) in a greater appreciation for the whole. More specifically, the mechanism of rainbow formation by the scattering of light from a transparent sphere is examined from a ray-theoretic viewpoint, for both homogeneous and radially inhomogeneous spheres. By examining the complementary approach of wave scattering theory, the resulting radial equations (for scalar and vector wave equations) can be regarded as time-independent Schrödinger-like equations. Consequently it is possible to exploit some of the mathematical techniques in potential scattering theory because every refractive index profile $n(r)$ defines a (wavenumber-dependent) scattering potential $V(k;r)$ for the problem. This is significantly different from the case of time-independent potential scattering in quantum mechanics because it ensures that there are no bound states of the system (this result is established in Appendix 2). The close correspondence between the resonant modes in scattering by a potential of the 'well-barrier' type and the behavior of electromagnetic 'rays' in a transparent (or dielectric) sphere is discussed in some detail.

**Acknowledgment**

I have been heavily influenced by the work of Professors H. M. Nussenzveig and J. A. Lock in the preparation of this Chapter. I would particularly like to thank Professor Lock for his generous advice, detailed and constructive suggestions on this material (also pointing out an error in Appendix 5), and for permission to use the quotation from his paper [3]. The comments of an anonymous reviewer also contributed significantly to the improvement of this Chapter, and are gratefully acknowledged.

**Appendix 1: The Debye series**

In [8, 13; see references therein] it is shown that, in terms of cylindrical Hankel functions of the first and second kinds,

$$\mathscr{S}_l(\lambda,\beta) = \frac{H_\lambda^{(2)}(\beta)}{H_\lambda^{(1)}(\beta)} R_{22}(\lambda,\beta) + T_{21}(\lambda,\beta) T_{12}(\lambda,\beta) \frac{H_\lambda^{(1)}(\alpha)}{H_\lambda^{(2)}(\alpha)} \sum_{p=1}^{\infty} \left[\rho(\lambda,\beta)\right]^{p-1} \quad (A1.1)$$

where

$$\rho(\lambda,\beta) = R_{11}(\lambda,\beta) \frac{H_\lambda^{(1)}(\alpha)}{H_\lambda^{(2)}(\alpha)}. \quad (A1.2)$$

This is the *Debye expansion*, arrived at by expanding the expression $\left[1-\rho(\lambda,\beta)\right]^{-1}$ as an infinite geometric series. The quantities $R_{22}, R_{11}, T_{21}$ and $T_{12}$ are respectively the external/internal reflection and internal/external transmission coefficients for the problem. This procedure transforms the interaction of 'wave + sphere' into a series of surface interactions. In so doing it 'unfolds' the stationary points of the integrand so that a given integral in the Poisson summation contains a few stationary points. This permits a ready identification of the many terms in accordance with ray theory. The first term represents direct reflection from the surface. The term $p=1$ has one such point (the transmitted ray), whereas $p=2$ has either two or zero stationary points (the former corresponding to the two supernumerary rays of the first-order rainbow). The $p^{th}$ term in the summation represents transmission into the sphere, via the term $T_{21}$ subsequently "bouncing" back and forth between $r=a$ and $r=0$ a total of $p$ times with $p-1$ internal reflections at the surface (this time via the $R_{11}$ term in $\rho$). The final factor in the second term $T_{12}$, corresponds to transmission to the outside medium. In general, therefore, the $p^{th}$ term of the Debye expansion represents the effect of $p+1$ surface interactions. Now $f(\beta,\theta)$ can be expressed as

$$f(\beta,\theta) = f_0(\beta,\theta) + \sum_{p=1}^{\infty} f_p(\beta,\theta), \quad (A1.3)$$

where

$$f_0(\beta,\theta) = \frac{i}{\beta} \sum_{m=-\infty}^{\infty} (-1)^m \int_0^\infty \left(1 - \frac{H_\lambda^{(2)}(\beta)}{H_\lambda^{(1)}(\beta)} R_{22}\right) P_{\lambda-1/2}(\cos\theta) e^{2\pi i m \lambda} \lambda \, d\lambda. \quad (A1.4)$$

This is the direct reflection term. The expression for $f_p(\beta,\theta)$ involves a similar type of integral for $p \geq 1$. The direct transmission term is the one of interest for zero-order bows, but the analysis of Nussenveig and co-workers deals with constant *n*, for which no such bow exists. As noted earlier, Lock [4] identified the existence of a zero bow for a Luneberg lens with focal length exceeding its radius. In general however, further work is necessary to determine the nature of direct transmission bows in other radially inhomogeneous transparent (or dielectric) spheres [18].

Returning to the constant *n* case, the application of the modified Watson transform to the third term ($p = 2$) in the Debye expansion of the scattering amplitude shows that it is *this term* which is associated with the phenomena of the primary rainbow. More generally, for a Debye term of given order *p*, a rainbow is characterized in the $\lambda$-plane by the occurrence of two real saddle points $\lambda$ and $\lambda'$ between *0* and $\beta$ in some domain of scattering angles *θ*, corresponding to the two scattered rays on the illuminated side. As $\theta \to \theta_R^+$ ($\theta_R$ being the *rainbow angle*) the two saddle points move toward each other along the real axis (Figure 7), merging together at $\theta = \theta_R$. As *θ* moves into the dark side, the two saddle points become complex, moving away from the real axis in complex conjugate directions. Therefore, as noted earlier, from a mathematical point of view a rainbow can be defined as a collision of two saddle points in the complex angular momentum plane. The primary bow light/shadow transition region is thus associated physically with the confluence of a pair of geometrical rays and their transformation into "complex rays".

**Appendix 2: Radially inhomogeneous media**

In electromagnetic scattering, for radially symmetric media, the electric field vector **E** must satisfy the scattering boundary conditions and the vector wave equation

$$\nabla \times \nabla \times \boldsymbol{E} - k^2 n^2(r) \boldsymbol{E} = \boldsymbol{0}. \quad \text{(A2.1)}$$

By expanding **E** in terms of vector spherical harmonics, the following radial equations are obtained for the *transverse electric* (*TE*) and *transverse magnetic* (*TM*) modes respectively [24]:

$$\frac{d^2 S_l(r)}{dr^2} + \left[ k^2 n^2(r) - \frac{l(l+1)}{r^2} \right] S_l(r) = 0; \quad \text{(A2.2)}$$

$$\frac{d^2 T_l(r)}{dr^2} - \frac{2n'(r)}{n(r)} \frac{dT_l(r)}{dr} + \left[ k^2 n^2(r) - \frac{l(l+1)}{r^2} \right] T_l(r) = 0. \quad \text{(A2.3)}$$

Each of these equations can be reworked into a time-independent Schrödinger equation form, with $\psi(r)$ now being a generic dependent variable for the two modes above. Thus:

$$\frac{d^2 \psi(r)}{dr^2} + \left[ k^2 - V(r) - \frac{l(l+1)}{r^2} \right] \psi(r) = 0, \quad \text{(A2.4a)}$$

or equivalently, as indicated earlier,

$$\frac{d^2\psi(r)}{dr^2} + \left[k^2 - V(r) - \frac{\lambda^2 - 1/4}{r^2}\right]\psi(r) = 0, \quad (A2.4b)$$

where $k^2 = E$ is the energy of the 'particle', $\lambda = l + 1/2$. The 'scattering potential' is now

$$V(r) = k^2 \left[1 - n^2(r)\right] \quad (A2.5)$$

for the *TE* mode, and (by eliminating the first derivative term in (A2.3); see (A2.9) below)

$$V(r) = k^2 \left[1 - n^2(r) + k^{-2} n(r) \frac{d^2}{dr^2}(n(r))^{-1}\right] \quad (A2.6)$$

for the *TM* mode. Thus for scattering by a dielectric sphere, the corresponding potential has finite range. Note that for constant refractive index, these two equations are identical in form. We examine one property of the equations (A2.4a, b)) above in more detail. Although they are formally identical to the radial Schrödinger equation, there are important differences for both the scalar and the vector problems. Pure 'bound state' solutions, that is real, regular and square-integrable solutions corresponding to $k^2 < 0$ (Im $k > 0$) do not in general exist in the 'non-QM case'. To see this, assume that $S_l(r)$ is a square-integrable solution of equation (A2.2). On multiplying by $\bar{S}_l(r)$ (the complex conjugate of $S_l(r)$) and integrating by parts, we obtain

$$\bar{S}_l(r) S_l'(r) \Big|_0^\infty - \int_0^\infty \left[|S_l'(r)|^2 + \left\{\frac{l(l+1)}{r^2} - k^2 n^2(r)\right\}|S_l(r)|^2\right] dr = 0. \quad (A2.7)$$

The integrated term vanishes because to be square-integrable, $S(r)$ must vanish at infinity, and we have noted already that near the origin, $S_l(r) \sim r^{l+1}$. Hence

$$\int_0^\infty \left[|S_l'(r)|^2 + \frac{l(l+1)}{r^2}|S_l(r)|^2\right] dr = \int_0^\infty k^2 n^2(r) |S_l(r)|^2 dr. \quad (A2.8)$$

Clearly, this cannot be satisfied for $k^2 < 0$ unless $n^2(r) < 0$ in some interval or set of intervals. This actually 'opens the door' for some insight into properties of 'metamaterials' for which the refractive index may be pure imaginary [40]. Regarding the second of the two potentials (A2.6), if we write $T_l(r) = U_l(r) n(r)$ then from (A2.3b) $U_l(r)$ satisfies the equation

$$\frac{d^2 U_l(r)}{dr^2} + \left[k^2 n^2(r) - n(r)\frac{d^2}{dr^2}\left[\frac{1}{n(r)}\right] - \frac{l(l+1)}{r^2}\right] U_l(r) = 0. \quad (A2.9)$$

A similar procedure to that above yields the less useful form

$$\int_0^\infty \left[ |U_l{}'(r)|^2 + \left\{ \frac{l(l+1)}{r^2} + n(r)\frac{d^2}{dr^2}\left(\frac{1}{n(r)}\right) \right\} |S_l(r)|^2 \right] dr = \int_0^\infty k^2 n^2(r) |U_l(r)|^2 \, dr. \quad (A2.10)$$

Clearly, this expression places some conditions on the concavity of $n^{-1}(r)$, but with the Liouville transformation [41] $r \mapsto s : s = \int_0^r n^2(t) dt,$ and $U_l \mapsto W_l : W_l(s) = m(s) W_l(s),$ where $m(s) = n(r(s))$, it follows that

$$\frac{d^2 U_l(r)}{dr^2} = m^2(s) \left[ m(s) \frac{d^2 W_l(s)}{ds^2} - W_l(s) \frac{d^2 m(s)}{ds^2} \right], \quad (A2.11)$$

and

$$\text{and } \frac{d^2}{dr^2}\left(\frac{1}{n(r)}\right) = -m^2(s) \frac{d^2 m(s)}{ds^2}. \quad (A2.13)$$

Therefore equation (A2.9) simplifies to the form

$$\frac{d^2 W_l(s)}{ds^2} + \left[ \frac{k^2}{m^2(s)} - \frac{l(l+1)}{m^4(s) r^2(s)} \right] W_l(s) = 0. \quad (A2.13)$$

The transformation $r \mapsto s$ is monotonic (and linear for $r > 1$), and $s \sim r$ in the neighborhood of the origin, so the previous analysis carries over, and we can conclude that for $n^2 > 0$ no bound states are possible.

**Appendix 3: Connection with classical scattering**

In the theory of classical scattering of a non-relativistic projectile particle of mass $m$ by a central force with potential $V(r)$, the total deflection angle $\theta$ is given by [8, 42]

$$\theta = \pi - 2b \int_a^\infty \frac{dr}{r^2 \left[ 1 - b^2/r^2 - V(r)/E \right]^{1/2}}, \quad (A3.1)$$

where $b$ is the impact parameter, $a$ is the distance of closest approach and $E$ is the particle energy. The integral can be recast to the optical case (using equation (9)) by setting $b = \sin i$ and

$$n(r) = \left[ 1 - \frac{V(r)}{E} \right]^{1/2}, \quad (A3.2)$$

with $V(r) < 0$ corresponding to an attractive potential with refractive index $n > 1$. This justifies the notion of a refracting sphere having the characteristics of a potential well, with implications, as we have noted, for morphology-dependent resonances.

**Appendix 4: The location of the *S*-matrix poles**

From equations (23) and (46), noting the implicit time-dependence $\exp(-i\omega t)$, we may write the asymptotic form of the solution for $\psi_l(r,t)$ as

$$\psi_l(r,t) = O\left(\frac{1}{r}\{e^{-ikr} - \mathscr{S}_l(k)e^{ikr}\}e^{-i\omega t}\right). \quad (A4.1)$$

The scattering matrix elements $\mathscr{S}_l(k)$ are given in terms of the Jost functions by equation (60), and since both functions $\tilde{f}_l(\pm k)$ are defined for complex values of $k$, (60) defines $\mathscr{S}_l(k)$ throughout the complex *k*-plane [43]. Using the probability conservation law (derived from the time-dependent Schrödinger equation)

$$\frac{\partial}{\partial t}\int_V |\psi|^2 dV = -\int_S \boldsymbol{j} \cdot d\boldsymbol{S}, \quad (A4.2)$$

where $\boldsymbol{j}$ is the probability flux density (in units for which $m = \hbar = 1$)

$$\boldsymbol{j} = \frac{i}{2}(\psi \nabla \bar{\psi} - \bar{\psi} \nabla \psi). \quad (A4.3)$$

The integration in (A4.2) is carried out on the surface of a large sphere of radius *R* such that the asymptotic solution (A4.1) may be used. Furthermore, if $\mathscr{S}_l(k)$ has a pole at the complex k-value $k = k_r + ik_i$, then the first term in (A4.1) may be neglected in the neighborhood of this point, and we may write

$$\psi_l(r,t) = \frac{u_l(r)}{r}e^{-i\omega t} = O\left(-\frac{\mathscr{S}_l(k)}{r}e^{i(kr-\omega t)}\right), \quad r \to \infty \quad (A4.4)$$

near the pole *k*. From (A4.2) we then find that

$$k_r k_i \int_0^R u_l^2(r) dr = -\frac{k_r}{2}|\mathscr{S}_l(k)|^2 e^{-2k_i R} < 0. \quad (A4.5)$$

Therefore it follows that either $k_r = 0$ (the poles of $\mathscr{S}_l(k)$ lie on the imaginary axis), or, if $k_r \neq 0$, then $k_i < 0$ (i.e. the poles of $\mathscr{S}_l(k)$ lie in the lower half-plane). Equivalently, the only poles in the upper half-plane must lie on the imaginary axis.

Note that in the above discussion, we have tacitly assumed that the angular frequency can be identified with the energy of the 'particle'. This is justified by virtue of the famous relation ' $E = h\nu \propto \omega$. Without loss of generality here we make set the constant of proportionality to be unity, whence $\omega = E = k^2$, so that

$$\omega = (k_r + ik_i)^2 = (k_r^2 - k_i^2) + 2ik_r k_i \equiv E_r - \frac{i\Gamma}{2}, \quad \Gamma = -4k_r k_i. \quad (A4.6)$$

**Appendix 5: Poles and resonances on the *k*-plane and *E*-plane.**

For algebraic simplicity, we consider the (simple) poles of the *S*-matrix for the one dimensional scalar problem [28, 44]. In this approach, the analysis is based on a slightly different formulation of the governing time-independent 'Schrödinger' equation, namely

$$\frac{1}{2}\frac{d^2 u(x)}{dx^2} + \left[k^2 - V(x)\right]u(x) = 0, \quad (A5.1)$$

For a square well of depth $V_0 > 0$ (i.e. $V(r) = -V_0$, $|x| < a/2$ and is zero elsewhere), the incident 'wave' is represented by

$$u(x) = Ae^{ikx}, \quad x < -a/2, \quad (A5.2)$$

and a transmitted wave

$$u(x) = Ae^{ik(x-a)} S(E), \quad x > a/2. \quad (A5.3)$$

The transmission coefficient $S(E)$ is the one-dimensional scattering matrix in this problem. It can be shown that [44]

$$S(E) = \left\{\cos Ka - \frac{i}{2}\left(\frac{k}{K} + \frac{K}{k}\right)\sin Ka\right\}^{-1}, \quad (A5.4)$$

where now $k = \sqrt{2E}$ and $K = \sqrt{2(E + V_0)}$. Note the similarity of this expression with the denominator of the S-matrix in equation (36). The transmissivity of the well is defined as

$$T(E) = |S(E)|^2 = \left\{1 + \frac{V_0^2 \sin^2 Ka}{4E(E + V_0)}\right\}^{-1}. \quad (A5.5)$$

This expression has maxima equal to one whenever $\sin Ka = 0$, i.e. when $Ka = n\pi$, $n = 1, 2, 3, \ldots$ Equivalently, $E = n^2 \pi^2 / 2a^2 - V_0 > 0$. These maxima correspond to resonances – perfect transmission – in this system. The well contains an integral number of half wavelengths when this condition is satisfied.

We examine $S(E)$ as an analytic function of the energy $E$ in what follows. For $E > 0$, $0 < T(E) \le 1$. Therefore, poles of $T(E)$ (and $S(E)$) can only occur when $-V_0 < E < 0$. In fact $S(E)$ has a pole whenever

$$\cos Ka - \frac{i}{2}\left(\frac{k}{K} + \frac{K}{k}\right)\sin Ka = 0, \quad (A5.6)$$

i.e. when

$$\cot Ka = \frac{1}{2}\left(\frac{K}{k} - \frac{k}{K}\right). \quad (A5.7)$$

Furthermore, from the identity $2\cot 2\theta = (\cot\theta - \tan\theta)$ the solutions of (A5.7) can be recast in terms of odd and even parity bound state solutions, i.e.

$$K\cot\left(\frac{Ka}{2}\right) = ik, \text{ and } K\tan\left(\frac{Ka}{2}\right) = -ik. \quad (A5.8a, b)$$

(Again, notice the similarity of (A5.8a) with $\alpha\cot\alpha = i\beta$ from equation (36)) Suppose now that a resonance occurs at $E = E_r \equiv k_r^2/2 > 0$. In the *vicinity* of such value of the resonance energy, we may expand the expression $\left(\frac{k}{K} + \frac{K}{k}\right)\tan Ka$ as

$$\left(\frac{k}{K} + \frac{K}{k}\right)\tan Ka = \frac{d}{dE}\left[\left(\frac{k}{K} + \frac{K}{k}\right)\tan Ka\right]_{E_r}(E - E_r) + O(E - E_r)^2. \quad (A5.9)$$

To first order in $(E - E_r)$, on simplifying, we find that

$$\left(\frac{k}{K} + \frac{K}{k}\right)\tan Ka \approx a\left[\frac{dK}{dE}\left(\frac{k}{K} + \frac{K}{k}\right)\right]_{E_r}(E - E_r) \equiv \frac{4}{\Gamma}(E - E_r). \quad (A5.10)$$

We can rewriting equation (A5.4) as

$$S(E) = \sec Ka \left\{1 - \frac{i}{2}\left(\frac{k}{K} + \frac{K}{k}\right)\tan Ka\right\}^{-1} \approx \sec Ka \left\{1 - i\frac{2}{\Gamma}(E - E_r)\right\}^{-1}$$
$$= \sec Ka \left(\frac{i\Gamma/2}{E - E_r + i\Gamma/2}\right) \approx \left(\frac{i\Gamma/2}{E - E_r + i\Gamma/2}\right). \quad (A5.11)$$

To this order of approximation, then, the pole of $S(E)$ lies in the fourth quadrant of the complex $E$-plane. There is a branch cut along the real axis, $E > 0$ since if $E = |E|e^{i\theta}$, and $E^{1/2} = |E|^{1/2} e^{i\theta/2}$, in the limit $\theta \to 2\pi^{-}$, $\sqrt{E} = -|E|^{1/2}$, and for $E < 0$, $k = i|2E|^{1/2}$. As can be seen from the term $\exp(ikx)$ in equation (A5.3), therefore, $E < 0$ corresponds to a decaying transmitted wave, and (A5.1) then defines the conditions for the bound states to exist within the potential well. These conditions are exactly the equations (A5.8a, b) above.

Similarly, for the more general three-dimensional case we would expect that, near a resonance, $\mathscr{S}_l(E)$ also has a pole in the fourth quadrant. This pole is in the analytic continuation of $\mathscr{S}_l(E)$ from above to below the positive real axis, and lies on the second Riemann sheet of $\mathscr{S}_l(E)$. The bound states of the well correspond to poles of $\mathscr{S}_l(E)$ on the negative real energy axis. The closer the resonances are to the real axis, the 'stronger' they become, that is, the more they behave like very long lived bound states [44].

Finally, a nice connection can be made to the phase shift from equation (A5.5). Retaining $E$ as the independent variable, we can write

$$S(E) = e^{i\delta(E)} |T(E)|^{1/2}. \quad (A5.12)$$

For notational convenience, we write equation (A5.4) as $S(E) = \left[A(E) - iB(E)\right]^{-1}$, with obvious choices for $A$ and $B$. Then it follows that

$$\tan \delta(E) = \frac{B(E)}{A(E)} = \frac{1}{2}\left(\frac{k}{K} + \frac{K}{k}\right)\tan Ka \approx \frac{2}{\Gamma}(E - E_r) \quad (A5.13)$$

on using equation (A5.10). Hence

$$\delta(E) \approx \arctan\left[\frac{2}{\Gamma}(E - E_r)\right]. \quad (A5.14)$$

Note also that

$$\frac{d\delta(E)}{dE} = \frac{2\Gamma}{\Gamma^2 + 4(E - E_r)^2}, \quad (A5.15)$$

And this derivative has a maximum value when $E = E_r$, that is at a resonance, so $\delta(E)$ varies rapidly there.

___

**[Figures (1 – 8) below]**

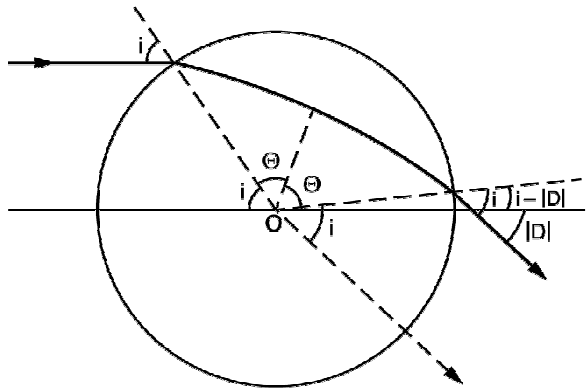

**Figure 1:** The ray path for direct transmission through a radially inhomogeneous sphere for $n'(r) < 0$.
___

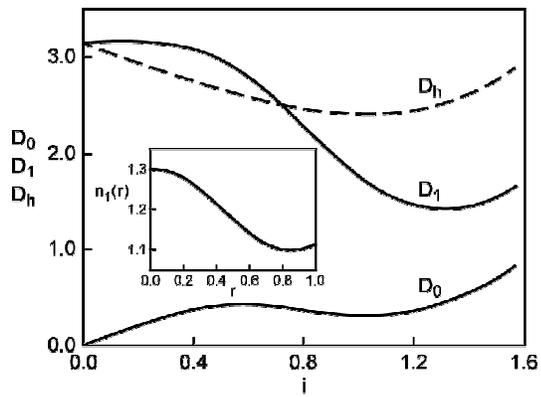

**Figure 2:** Deviation functions for both a homogeneous ($D_h$) and inhomogeneous spheres ($D_0$ and $D_1$) for the profile $n_1(r)$ [inset].

___

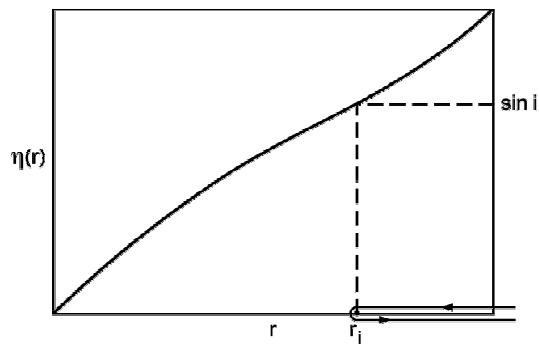

**Figure 3:** $\eta(r) = rn(r)$ for the monotonic case. The point of closest approach is $r = r_c$.

___

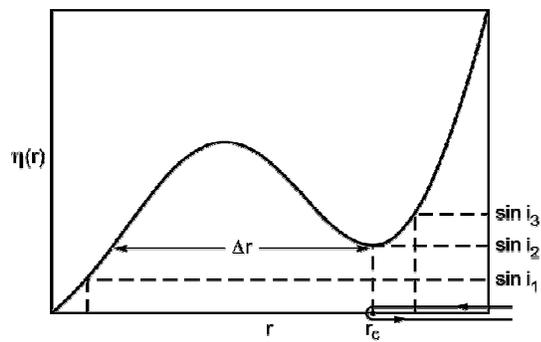

**Figure 4:** $\eta(r) = rn(r)$ for the non-monotonic case. The point of closest approach for $i > i_2$ is $r = r_c$, and a zone of width $\Delta r$ exists into which no ray penetrates.

________________________________________________________________________

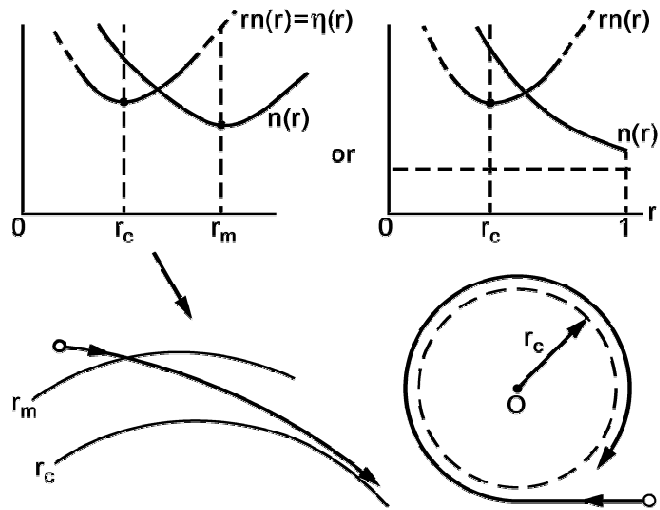

**Figure 5:** The phenomenon of orbiting illustrated schematically associated with a zero of $\eta'(r)$ showing the $\eta(r)$ and $n(r)$ profiles associated with the existence of a 'critical' ray separating two types of ray behavior (upper diagrams). The lower diagrams illustrate two different ways in which rays can approach the critical radius $r_c$. (See equation (14) and the associated discussion in section 2.2)

________________________________________________________________________

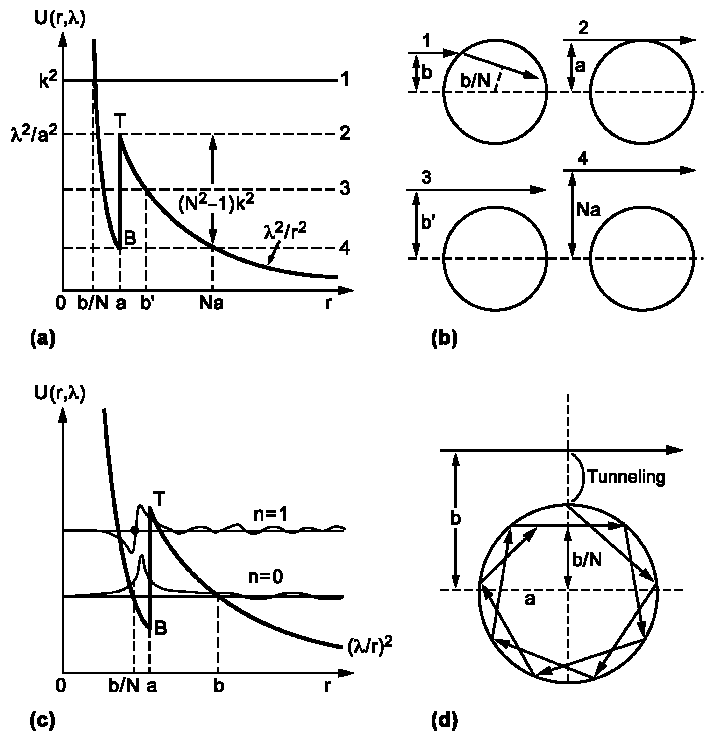

**Figure 6(a–d)** (redrawn from [8])**:**
(a) The effective potential $U(r)$ for a transparent sphere of radius $a$ showing four `energy levels', respectively above the top of the potential well, at the top, in the middle and at the bottom of the well. Note that the constant refractive $n$ has temporarily been replaced by $N$ to distinguish it from the node number $n$ in (c).
(b) The corresponding incident rays and impact parameters. Case 2 shows a tangentially-incident ray; note that in Case 1 the refracted ray is shown. It passes the center at a distance of $l = b/N$; that this is the case is readily shown from simple geometry: from Snel's law of refraction $\sin i = N \sin r = b/a$, and since $l = a \sin r$, the result follows directly.
(c) Similar to (a), but with resonant wave functions shown, corresponding to node numbers $n = 0$ and $n = 1$ (the latter possessing a single node).
(d) The `tunneling' phenomenon illustrated for an impact parameter $b > a$, being multiply reflected after tunneling, between the surface $r = a$ and the caustic surface $r = b/N$ (the inner turning point).

______________________________________________________________________________

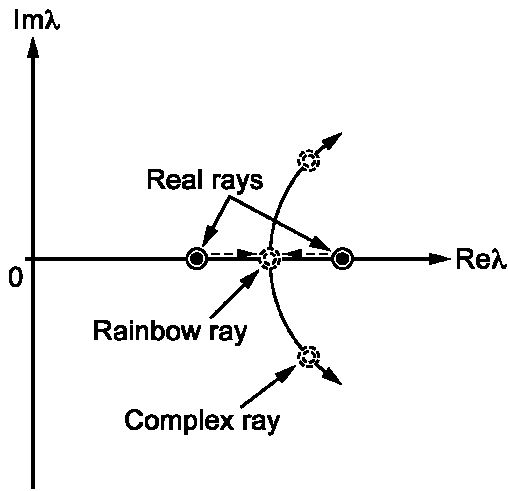

**Figure 7** (redrawn from [11]): The 'collision of two real saddle points in the complex λ-plane as the rainbow angle $(\theta_R)$ is approached from below (i.e. from the illuminated side). At $\theta_R$ the points collide and subsequently move away from each other along complex-conjugate directions as $\theta$ increases away from $\theta_R$ into the shadow region. It is the lower complex saddle-point that contributes to the wave field in this region.

___

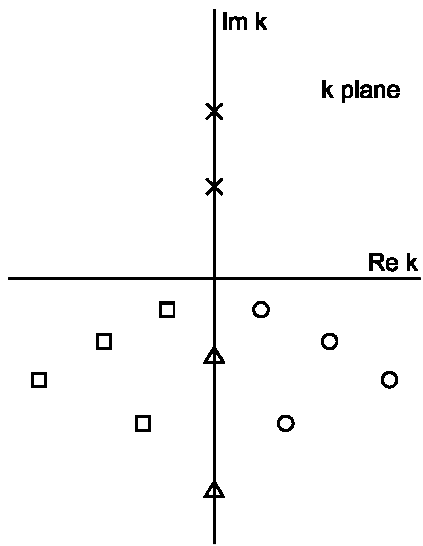

**Figure 8** (redrawn from [27a]): A generic distribution of poles for the *S*-matrix. Crosses correspond to bound state poles; circles to resonance poles; squares to their conjugate poles, and triangles to virtual states.

___

Page 43 of 43